\UseRawInputEncoding
\documentclass[aip,pop,reprint]{revtex4-2}
\usepackage{dcolumn}
\usepackage{amsmath}
\usepackage{bm}
\usepackage{hyperref}
\usepackage{natbib}
\usepackage{amsfonts}
\usepackage{booktabs}
\usepackage{siunitx}
\usepackage{relsize}
\usepackage{mathtools}
\usepackage[colorinlistoftodos]{todonotes}
\usepackage{physics}
\usepackage{graphicx}
\usepackage{array}
\usepackage{float}
\usepackage{svg}
\usepackage{tikz}
\usetikzlibrary{tikzmark}
\usepackage{empheq}
\usepackage{enumitem}
\makeatletter
\def\namedlabel#1#2{\begingroup
    #2%
    \def\@currentlabel{#2}%
    \phantomsection\label{#1}\endgroup
}

 \usepackage{caption}
\makeatother
\usepackage{tcolorbox}
\usepackage{color}
\usepackage{amssymb}

\begin{document}

\title{Strong Coulomb Coupling Influences Ion and Neutral Temperatures in Atmospheric Pressure Plasmas}
\author{M. D. Acciarri$^1$, C. Moore$^2$, S. D. Baalrud$^1$}
\address{\textsuperscript{1} Department of Nuclear Engineering and Radiological Sciences, University of Michigan, Ann Arbor, MI 48109, USA}
\address{\textsuperscript{2} Sandia National Laboratories, Albuquerque, NM 87185, USA}


\date{\today}

\begin{abstract}

Molecular dynamics simulations are used to model ion and neutral temperature evolution in partially-ionized atmospheric pressure plasma at different ionization fractions. Results show that ion-ion interactions are strongly coupled at ionization fractions as low as $10^{-5}$ and that the temperature evolution is influenced by effects associated with the strong coupling. Specifically, disorder-induced heating is found to rapidly heat ions on a timescale of the ion plasma period ($\sim10$s ps) after an ionization pulse. This is followed by the collisional relaxation of ions and neutrals, which cools ions and heats neutrals on a longer ($\sim$ns) timescale. Slight heating then occurs over a much longer ($\sim 100$s ns) timescale due to ion-neutral three-body recombination. An analytic model of the temperature evolution is developed that agrees with the simulation results. 
A conclusion is that strong coupling effects are important in atmospheric pressure plasmas. 

\end{abstract}


\maketitle





\section{Introduction}

Atmospheric pressure plasmas, in particular cold atmospheric pressure plasmas (CAPP), have been widely tested for numerous applications including the inactivation of various pathogens in medicine \cite{app11114809}, food industry \cite{https://doi.org/10.1002/ppap.201700085}, agriculture \cite{misra_cold_2016}, and water purification \cite{Adamovich_2022}, among others, showing promising results and an increasing interest in those industries. CAPP technology presents several advantages over other plasma technologies including operational simplicity, low running cost, and environmental friendliness, since no vacuum chamber is needed and the reactor is frequently open \cite{app11114809}. The increasing interest shown in atmospheric pressure plasmas and the lack of understanding about the main mechanisms involved in the plasma dynamics, transport of reactive species and chemistry motivates a better understanding of the physics involved in order to improve the development of plasma sources that promote reactions of interest and make devices more efficient \cite{Neyts_2014,Bogaerts_2020}. 
Here, we show an example of such a fundamental physics effect: ion-ion interactions are strongly coupled and the ion and neutral temperatures are influenced by an associated disorder-induced heating (DIH) process. 

Strong coupling refers to interactions in which the average potential energy of interacting particles [$\phi_{ss^\prime}(r=a_{ss^\prime})$, where $a_{ss^\prime}$ is the average distance between particles of species $s$ and $s^\prime$] exceeds their average kinetic energy ($k_BT_{ss'}$), i.e. $\Gamma_{ss^\prime} \gtrsim 1$, where $s$ and $s'$ are the species involved in the interaction and,
\begin{equation}
    \centering
    \Gamma_{ss'} = \frac{\phi_{ss'}(r=a_{ss^\prime})}{k_B T_{ss^\prime}}.
    \label{eq:gamma}
\end{equation} 
For example, the Coulomb potential for ion-ion interactions is $\phi_{ii}=(Ze)^2/4\pi \epsilon_0 a_{ii}$ and the ion-ion Coulomb coupling parameter is $\Gamma_{ii} = (Ze)^2/(4 \pi\epsilon_0 a_{ii} k_B T_{i})$. 
In most CAPPs, ions are expected to be in equilibrium with the neutral gas near room temperature, while electrons have a much higher temperature on the order of eV or several eV. 
A consequence is that the hotter electrons are characterized by a weakly coupled regime ($\Gamma_{ee} \ll 1$), while the much cooler ions can be strongly coupled ($\Gamma_{ii} \gtrsim 1)$.

Strongly coupled plasmas are influenced by physical effects that fundamentally differ from those governing weakly coupled plasmas. 
An example is disorder-induced heating. 
This arises in strongly coupled systems when the potential energy landscape changes in such a way that particles move to a lower potential energy configuration, releasing kinetic energy in the form of heat. 
One way that this can occur is through the ionization of a neutral gas. 
When short-range atomic interactions change to long-range Coulomb interactions, ions reconfigure to a more ordered state due to their mutual repulsion. 
The more ordered state has a lower potential energy than the initial state (immediately after ionization) because ions spread further apart from one another. 
The decrease in potential energy in this reconfiguration is compensated by an increase in kinetic energy. 
Disorder-induced heating is not important in weakly coupled plasmas because the kinetic energy gained in the reconfiguration is small compared to the initial kinetic energy. 
However, it can be a dominant effect in strongly coupled plasmas. 
For example, DIH caused by ionization has been measured and studied in detail in ultracold neutral plasmas \cite{doi:10.1063/1.3366240,Pohl_Pattard_Rost,DIH_UCNP,Kuzmin_ONeil}. 

Here, we show that ionization can cause DIH in atmospheric pressure plasmas as well, and that it can be a dominant effect that determines both the ion and neutral temperature evolution. 
Using Molecular Dynamics (MD) simulations, we show that immediately after ionization ions have a coupling strength much larger than one, with $\Gamma_{ii}$ ranging from approximately 5 to 25 for simulations with ionization fractions ranging from 0.01 to 0.7. 
Ions then rapidly heat over a timescale of approximately one ion plasma period ($\omega_{pi}^{-1}$, where $\omega_{pi} = \sqrt{e^2 n_i/\epsilon_o m_i}$ is the ion plasma frequency) to a condition where $\Gamma_{ii} \approx 1$. 
This typically corresponds to a $\sim$ps timescale. 
This can raise the ion temperature several times, depending on the ionization fraction. 
After DIH, ions thermally equilibrate with neutrals via elastic scattering, causing them to cool and the neutral gas to heat. 
The cooling causes ions to return to a more strongly coupled state. 
Depending on the ionization fraction, the neutral heating can be substantial, leading to neutral temperatures that are several times room temperature. 
Ion-neutral thermal equilibration typically takes hundreds to thousands of ion plasma periods, corresponding to a $\sim$ns timescale. 
In parallel with these comparatively fast processes, a weaker heating of both ions and neutrals is observed over a much longer timescale ($\sim100$s ns) due to ion-neutral three-body recombination. 
In addition to the simulations, a model is developed to describe the main features of the ion and neutral temperature evolution.  


An implication of these findings is that common approaches to modeling atmospheric pressure plasmas need to be reassessed. 
Most modeling techniques, including Particle-In-Cell (PIC) methods and solutions of multi-fluid equations, are based on solving a Boltzmann equation or approximations of it (such as multi-fluid models obtained from taking moments of the Boltzmann equation). 
However, the Boltzmann equation is valid only for weakly coupled systems, such as dilute gases and plasmas. 
At a fundamental level, it does not treat strong coupling effects that can influence dynamics of atmospheric pressure plasmas. 
Disorder-induced heating is an example of this. 
If strong correlations are not accounted for, the ion and neutral temperatures could be mistakenly underestimated. 

Although ion temperatures are difficult to measure in atmospheric pressure plasmas, measurements have been made of neutral gas temperatures that are elevated well above the ambient room temperature conditions \cite{Pai_2010,Rusterholtz_2013,Popov_2011,Popov_2016,Ono_2008,Mintoussov_2011,HofmannPSST2011}. 
Some explanations for this have been postulated. 
One is that energy transfer from electron-neutral elastic collisions heats neutrals \cite{vanderHorst2012}. 
Although some heating through this mechanism is certain, it is not clear if it happens at a fast enough rate or if the resulting ion temperature is correct to explain the measurements. 
Other possible explanations for a more rapid transfer of energy from electrons to neutral atoms/molecules include energy transfer between vibrationally excited states in molecules \cite{Adamovich2000,Guerra_2019}, energy transfer from vibrational to translational states \cite{Adamovich2000,Guerra_2019,Ono_2008}, electron-ion recombination processes \cite{Mintoussov_2011} followed by ion-neutral elastic collisions, electron impact dissociation reactions \cite{Popov_2011} and quenching of electronically excited molecules by oxygen atoms \cite{Popov_2016}. 
Although each is a plausible mechanism, there hasn't been a sufficiently detailed study to test if they explain the experimental measurements. 
Here, we propose that DIH is another mechanism that should be considered. 
We compare the predicted neutral gas temperature after ions and neutrals have thermally equilibrated to an experiment \cite{vanderHorst2012}, and find that the predictions are close to the measured temperature. 
These results are relevant for a number of plasma sources that achieve high ionization fractions at atmospheric and elevated pressures, including microdischarges~\cite{DongAPL2005}, spark filaments~\cite{ParkevichPSST2019,MinesiPSST2020,BatallerPRL2016}, laser-produced plasmas~\cite{BatallerPRL2014,BatallerOL2019}, arc filaments~\cite{Verreycken_2012} and the anode region of electric arcs~\cite{Heberlein_2009}, among others. 

In the next section the coupling parameter associated with ion-ion, ion-neutral and neutral-neutral interactions is shown in a pressure--ionization fraction space, predicting that at atmospheric pressure and over a wide range of ionization fractions ions are expected to be strongly coupled. In section \ref{sec:MD_simulations} the MD simulation setup is described. 
This is followed by a discussion of the ion and neutral temperature evolution in section \ref{sec:MD_results}. 
In section \ref{sec:model} a theoretical model for predicting the maximum and equilibrium ion temperature as a function of the ionization fraction and pressure is described. Finally, section \ref{sec:comparison} shows a comparison of the observed heating in experiments conducted at atmospheric pressure and the prediction for the temperature using our model.

\section{Coupling Parameter Space} \label{sec:coupling_space}

\begin{figure}
    \centering
    \includegraphics[width=\linewidth]{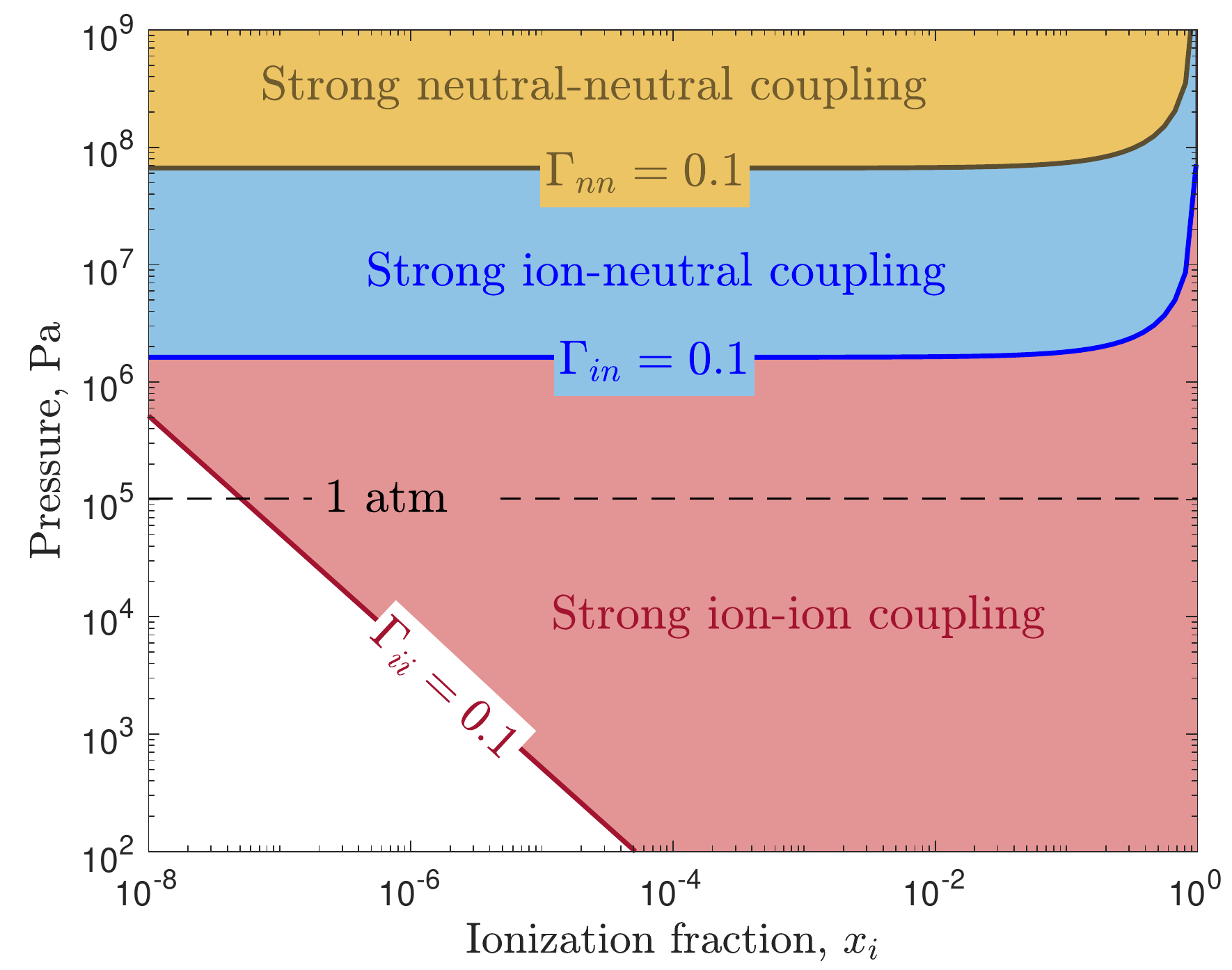}
    \caption{Coupling parameter space for ion-ion, ion-neutral and neutral-neutral interactions for an Ar gas at room temperature and  different pressures and ionization fractions.}
    \label{fig:gamma_diagram}
\end{figure}
For a partially ionized plasma with one gas species, neutral-neutral, ion-neutral and ion-ion interactions are modeled using the Lennard-Jones, charge induced dipole and Coulomb potentials respectively~\cite{doi:https://doi.org/10.1002/0471724254.ch3} 
\begin{equation}
    \phi_\textrm{LJ}(r) = 4\epsilon\left[ \left(\frac{\sigma}{r}\right)^{12} - \left(\frac{\sigma}{r}\right)^{6}\right] ,
    \label{eq:LJ}
\end{equation}
\begin{equation}
    \phi_\textrm{ind}(r) = - \frac{q^2}{8 \pi \epsilon_0} \frac{\alpha_\textrm{R} a^3_0}{r^4} ,
    \label{eq:induced}
\end{equation}
and
\begin{equation}
    \phi(r) =  \frac{q^2}{4 \pi \epsilon_0} \frac{1}{r} ,
    \label{eq:coulomb}
\end{equation} 
where $\epsilon=120 k_B$, $\sigma=0.34$ nm, and $\alpha_\textrm{R}=11.08$ for Ar \cite{doi:https://doi.org/10.1002/0471724254.ch3} and $a_0$ is the Bohr radius. Considering an Ar plasma at room temperature and variable ionization fraction and pressure, the coupling parameter associated with each interaction can be computed from equations~(\ref{eq:gamma})--(\ref{eq:coulomb}), using $a_{in} = (3/4\pi n_{in})^{1/3}$ where $n_{in}\approx x_i n_i+x_n n_n$ to estimate the average interparticle spacing between ions and neutrals. 
Here, $x_i = n_i/n$ and $x_n = n_n/n$ are the ion and neutral concentrations, $n_i$ and $n_n$ are the ion and neutral densities, and $n= n_i + n_n$ is the total density.  
The pressure and ionization fraction at which the transition from a weakly to a strongly coupled regime occurs is estimated from the limit $\Gamma = 0.1$, which is the condition where the Boltzmann equation is known to break down \cite{Scheiner_2020,Daligault_diffusion}. 

As shown in figure \ref{fig:gamma_diagram}, at small ionization fractions and pressures below atmospheric pressure, none of the interactions are strongly coupled. However, by increasing the ionization fraction or the pressure, it is possible to find a strong ion-ion coupling region. 
In fact, the figure shows that ion-ion interactions at many CAPP conditions are expected to be strongly coupled. If the pressure is increased above $\approx$ 10 atm, a strong ion-neutral coupling regime is reached. Further increases in pressure can lead to a strong neutral-neutral coupling regime at around 1000 atm.

Since the ion-ion Coulomb coupling parameter exceeds 0.1 in many CAPPs applications, the binary ion-ion collision picture and hence the Boltzmann equation are not expected to apply.
In this strongly coupled regime, many-body collisions dominate the interactions between ions and a first-principles approach is required. Hence, we used MD simulations, which provide a first-principles solution of Newton's equation of motion for the dynamics of $N$ interacting particles. 

\section{Molecular Dynamics Simulations} \label{sec:MD_simulations}


Molecular dynamics simulations were carried out using the open-source software LAMMPS \cite{LAMMPS}. Since electrons are much hotter than ions and are weakly coupled, they are treated as a background non-interacting species when modeling ion and neutral dynamics. 
Thus, they were not included in the simulation. 
This is similar to the one-component plasma model~\cite{BausPR1980}, which is known to provide an accurate description of ions in the presence of weakly coupled (comparatively hot) electrons, such as in ultracold neutral plasmas~\cite{KILLIAN200777}. 
Short (neutral-neutral), medium (ion-neutral) and long (ion-ion) range interactions were modeled using the potentials defined in equations (\ref{eq:LJ})--(\ref{eq:coulomb}).

Since the charge induced dipole potential is attractive, particles can interact at short spatial scales. 
In order to avoid the rare occurrence of close interactions that require a very short timestep to resolve, a repulsive core term was added to the charge-induced dipole potential from expression~(\ref{eq:induced})
\begin{equation}
    \centering
    \phi_\textrm{ind}(r) = \frac{q^2}{8 \pi \epsilon_0} \frac{\alpha_R a^3_0}{r^4} \biggl( \frac{r_{\phi}^8}{r^8} - 1 \biggr) ,
    \label{eq:induced_numerical}
\end{equation} 
where $r_{\phi}$ is the radius at which the repulsive core acts. 
It was desired to choose $r_{\phi}$ to be small enough to minimize the occurrence of non-physical force values, but large enough to decrease the computational cost due to the smaller timestep requirement. 
This is similar to what has been done in MD simulations of ultracold neutral plasmas~\cite{KuzminPRL2002,TiwariPRE2017}. 
The effect of the repulsive core also depends on the simulation setup. For simulations conducted at thermodynamic equilibrium, we found that the value of $r_{\phi}$ sets the number of ions that attach to neutrals; as shown in the Appendix \ref{sec:A}. However, this work concentrates on non-equilibrium simulations where $r_{\phi}$ is only a numerical convergence parameter needed to avoid closely orbiting particles. The reason why $r_{\phi}$ is not important in the non-equilibrium simulations is that ion-neutral three-body recombination is slow compared to the DIH and ion-neutral relaxation timescales that we focus on. 
For the non-equilibrium simulations we chose $r_{\phi} = 0.133 a_\textrm{in}$ (here $a_\textrm{in}$ was estimated as $\approx a_\textrm{nn}$ for $x_i<0.5$ and $\approx a_\textrm{ii}$ for $x_i>0.5$). This is the smallest value at which we can avoid closely orbiting particles and non-physical force values; see Appendix \ref{sec:A} and figure \ref{fig:convergence_phi}. A convergence test was conducted varying  $r_{\phi}$ from 0.133$a_\textrm{in}$ to 0.5$a_\textrm{in}$ at an ionization fraction of $x_i=0.5$ (figure \ref{fig:convergence_phi}) showing that the ion and neutral temperature at the end of the simulation (after ion-neutral relaxation) is independent of $r_{\phi}$ when $r_{\phi}<0.25 a_\textrm{in}$.

\begin{figure}
    \centering
    \includegraphics[width=\linewidth]{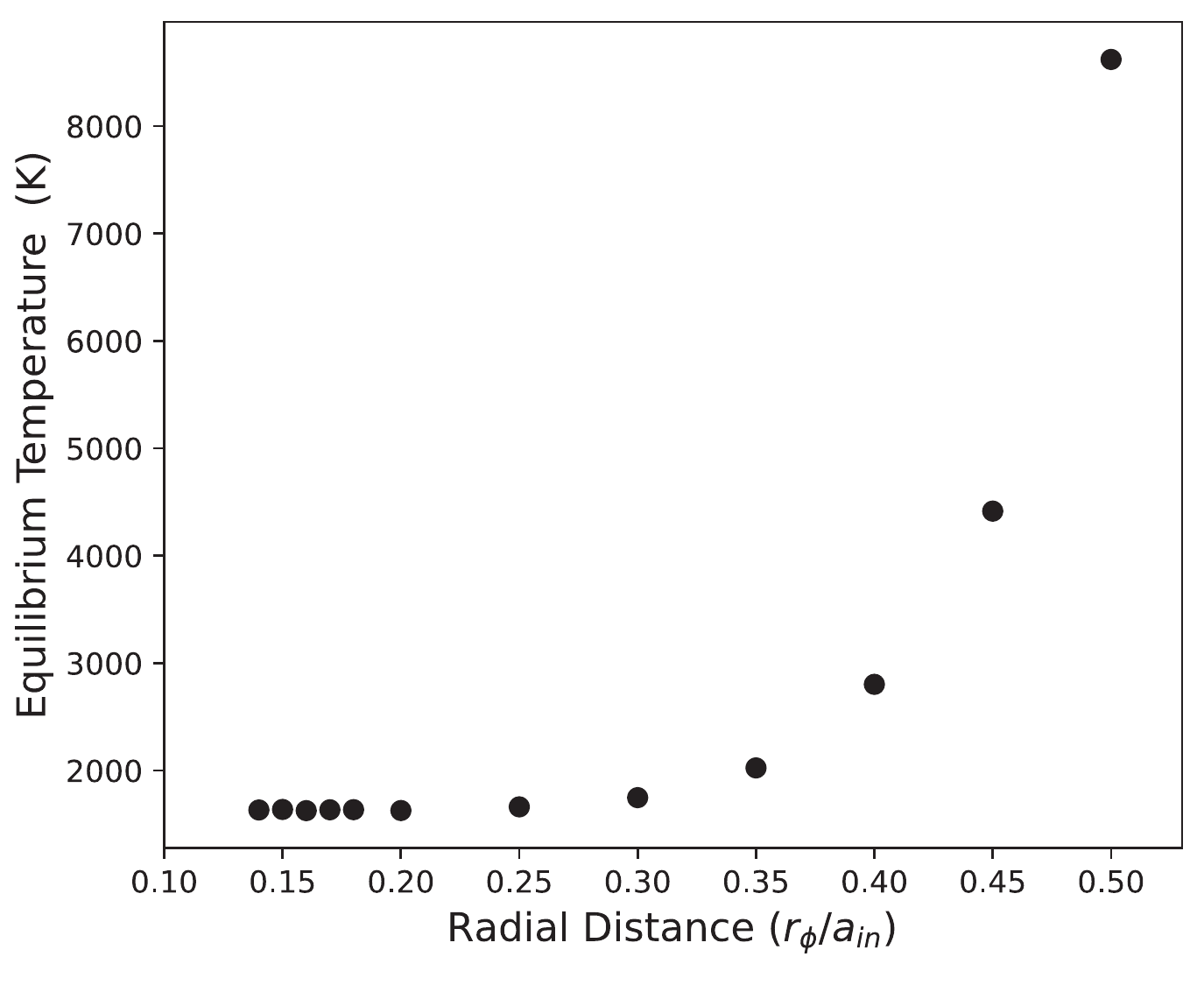}
    \caption{Ion and neutral final temperatures for different $r_{\phi}$ values.}
    \label{fig:convergence_phi}
\end{figure}


In order to study the evolution of a non-equilibrium discharge, a neutral Ar gas at room temperature and atmospheric pressure was simulated until equilibrium was reached. 
This stage of the simulation was run with a Nos\'{e}-Hoover thermostat (NVT ensemble) applied \cite{MD_Frenkel}. 
Then, a fraction of the particles were instantly ionized and a NVE (microcanonical) simulation was run including the ion-neutral, neutral-neutral and ion-ion interactions. This simulation setup was repeated for different ionization fractions. 
The timestep used was $5\times10^{-4}\omega_{pi}$ and the NVE simulation was run until the equilibrium was reached. The number of particles was varied such that the minimum number of ions in the system was 2500 and the number of neutral atoms was scaled in order to achieve the desired ionization fraction in each simulation. The simulation domain size was scaled with the total number of particles in order to maintain the desired gas density. The simulation domain was a three-dimensional box with periodic boundary conditions. The ionization fraction, total number of particles, time step and length of simulation used for each simulation are specified in table \ref{tab:simulation_parameters}. 

\begin{table}
\centering
\caption{Parameters used for the Molecular Dynamics simulations. The time step value was $\Delta t\times\omega_{pi} = 5\times10^{-4}$ in all cases. }
\label{tab:simulation_parameters}
\begin{tabular}{|c|c|c|c|}
\hline
\begin{tabular}[c]{@{}c@{}}Ionization \\ Fraction \\ ($x_i$) \end{tabular} & \begin{tabular}[c]{@{}c@{}}Number of\\ Particles\end{tabular} & \begin{tabular}[c]{@{}c@{}}Plasma \\ Frequency\\ ($\omega_{pi}$) \\ $\times$ $10^{12}$ \; (\textrm{rad/s}) \end{tabular} & \begin{tabular}[c]{@{}c@{}}Length of \\ Simulation \\  ($t_f$ $\times$ $\omega_{pi}$)\end{tabular} \\ \hline
0.01 & 250000 & 0.1046 &  150 \\ \hline
0.1 & 25000 & 0.3308 &  500 \\ \hline
0.3 & 10000 & 0.5731 &  650 \\ \hline
0.5 & 10000 & 0.7398 &  800\\ \hline
0.7 & 10000 & 0.8754 &  1000\\ \hline
\end{tabular}
\end{table}

\section{Molecular Dynamics Results} \label{sec:MD_results}

\begin{figure}[H]
    \centering
    \includegraphics[width=\linewidth]{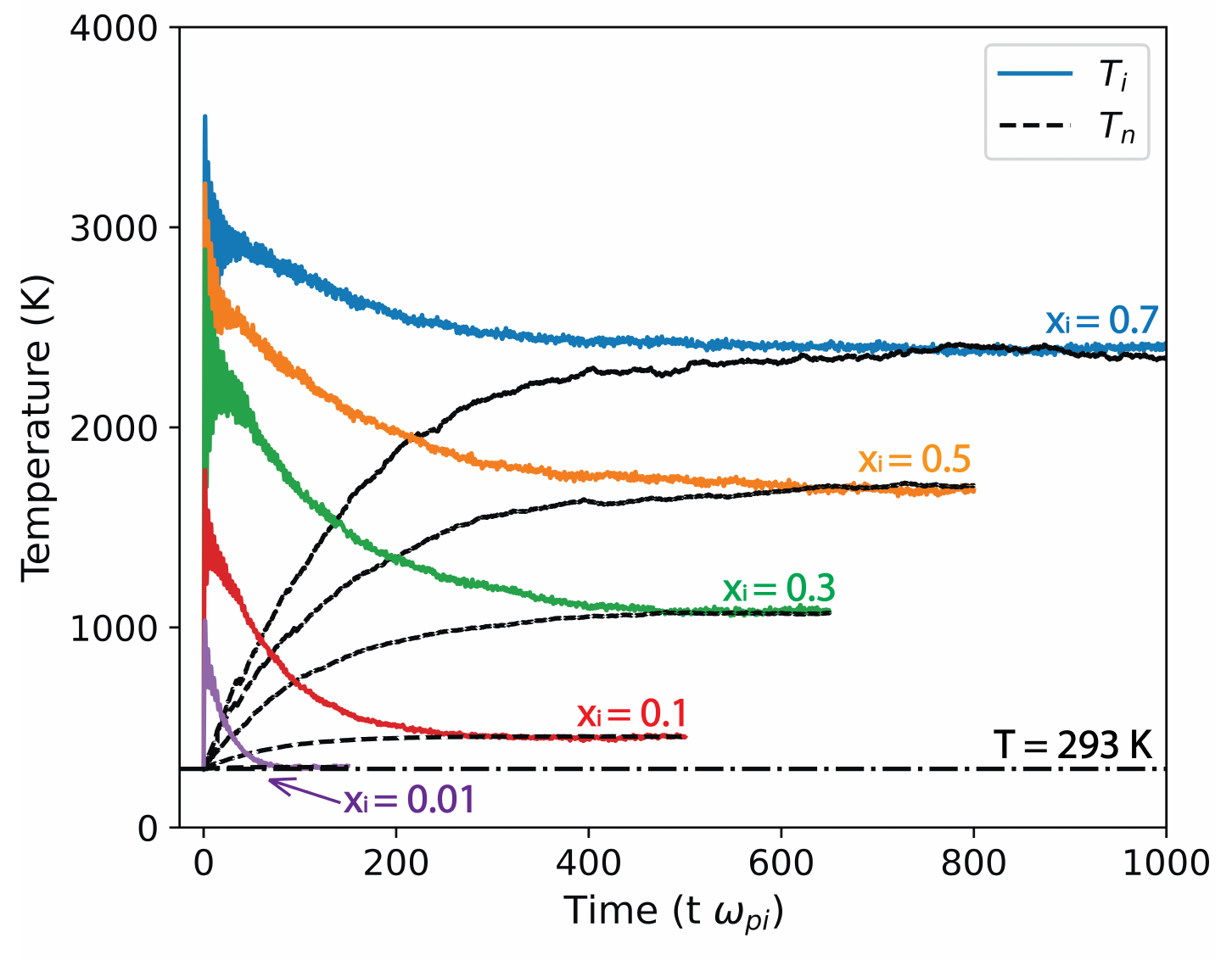}
    \caption{Evolution of the ion and neutral temperatures during the discharge at atmospheric pressure for different ionization fractions.}
    \label{fig:Ti_t_xi}
\end{figure}

As shown in figure~\ref{fig:Ti_t_xi}, the evolution of the ion temperature can be divided into three stages. First, a rapid increase in the ion temperature was observed over the first ion plasma period of the simulation. 
This is thought to be due to disorder induced heating.
This stage generates fluctuations in the ion temperature that persist for several plasma periods; as shown in figures~\ref{fig:Ti_t_xi} and \ref{fig:energy_fluctuations_xi_0_01}. Secondly, ion-neutral temperature relaxation was observed with a timescale  characterized by the ion-neutral collision frequency. 
Finally, a gradual heating of both ions and neutrals is observed over a much longer timescale due to three-body recombination of ions and neutrals (this is difficult to view in figure~\ref{fig:Ti_t_xi}, but is demonstrated more clearly below). 

\subsection{Disorder Induced Heating}

\begin{figure}
    \centering
    \includegraphics[width=\linewidth]{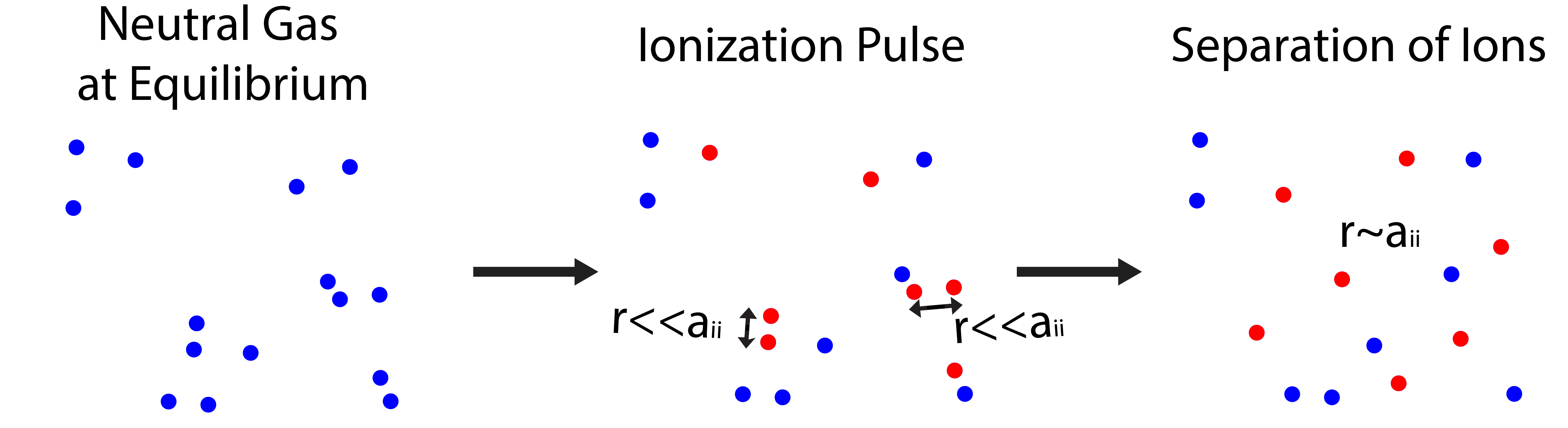}
    \caption{After an instant ionization pulse, ions can be at distances much smaller compared to $a_{ii}$ leading to large repulsion forces.}
    \label{fig:DIH_Figure}
\end{figure}

Before the ionization pulse is applied, the initial distribution of positions of neutral atoms corresponds to the equilibrium state of a gas interacting via the short-range Lennard-Jones potential. Therefore, ionization creates many ions separated by distances that are much smaller than the average distance between ions $a_{ii}=(3/4\pi n_i)^{1/3}$ where $n_i$ is the ion density. Since the interaction between ions is governed by the Coulomb potential, which is a long range potential, this leads to a large repulsion force between ion pairs that brings ions apart, as illustrated in figure~\ref{fig:DIH_Figure}. 

\begin{figure}
    \centering
    \includegraphics[width=\linewidth]{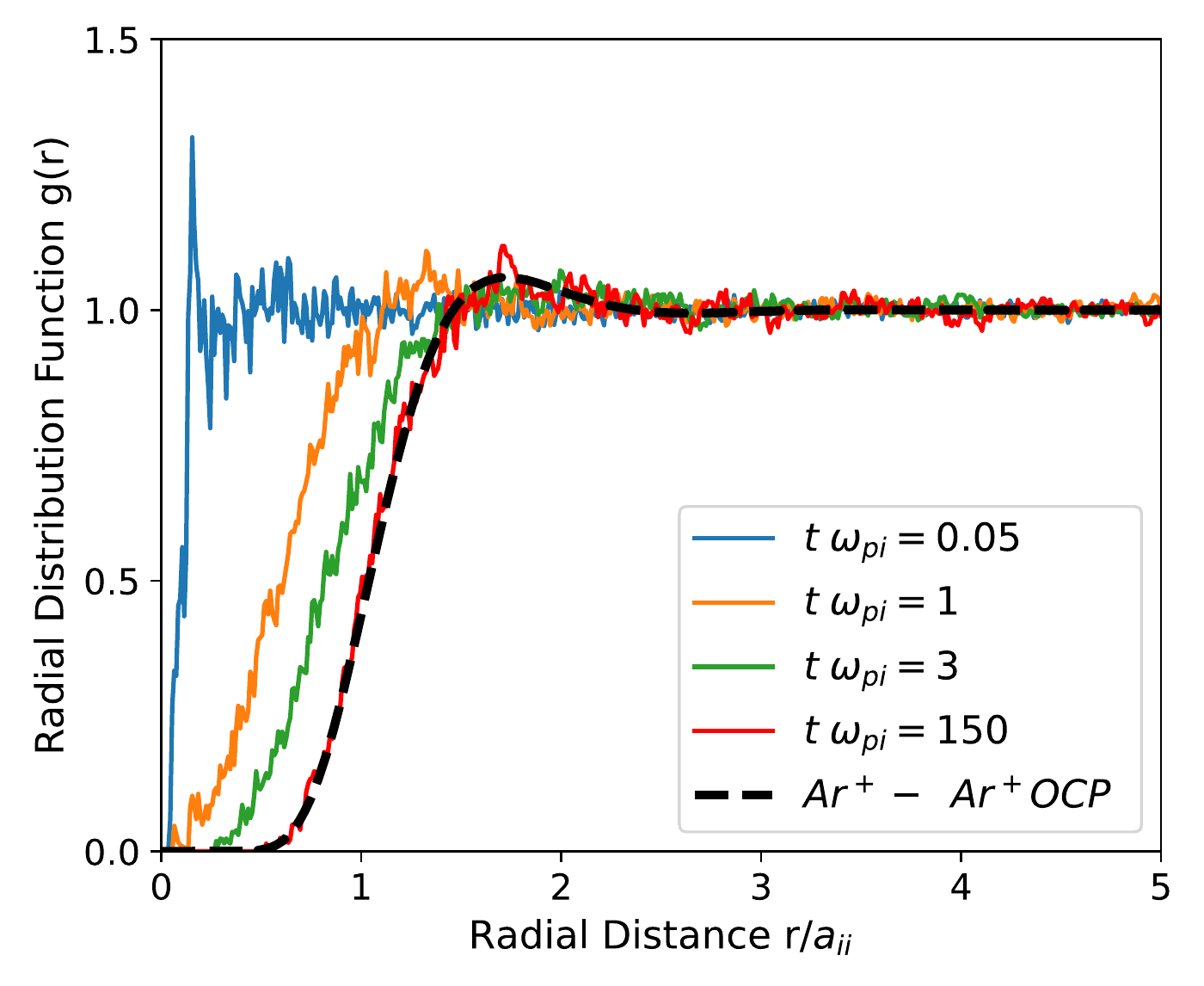}
    \caption{Ion-ion radial distribution function $g(r)$ at different timesteps during the simulation of a partially ionized Ar plasma with an ionization fraction of $x_i=0.01$. The ion-ion radial distribution function corresponds to an OCP at the same equilibrium $\Gamma_{ii}$ showing that the ions are strongly coupled and the ion-ion interactions are not screened by the presence of neutral atoms. In each simulation, the $g(r)$ was obtained after computing an average over all the ions in the simulation and 100 consecutive time steps that corresponded to a time window of $t\times\omega_{pi}=0.05$. An ensemble average was also computed using data from 20 different simulations.}
    \label{fig:gr_0_01}
\end{figure}

The separation of ions corresponds to the formation of a correlated state. 
This can be quantified by the radial distribution function $[g(r)]$, which is defined by setting $n_o g(r)4 \pi r^2 dr = N(r)$, where $N(r)$ is the total number of particles in a spherical shell of radius $r$ and thickness $dr$ centered on a chosen particle. 
Here, $n_o = N/V$ is the average number of particles ($N$) in a volume ($V$); i.e., it is the uniform background number density. 
When $g(r) =1$ for all distances $r$, the system is in an uncorrelated state. 
Figure~\ref{fig:gr_0_01} shows that just after ionization the ion positions correspond to the weakly correlated state of the neutral gas they were formed from. 
Correlations quickly develop as ions spread apart over the timescale of an ion plasma period. 
This is indicated by the void of particles that forms from distances $r=0$ to approximately $r=a_{ii}$. 
Such a void, which is sometimes referred to a Coulomb hole in plasmas, is a characteristic property of a strongly coupled system. 
Another characteristic is a peak near the average nearest neighbor distance ($a_{ii}$), which is also observed at times later than one ion plasma period. 

\begin{figure}
    \centering
    \includegraphics[width=\linewidth]{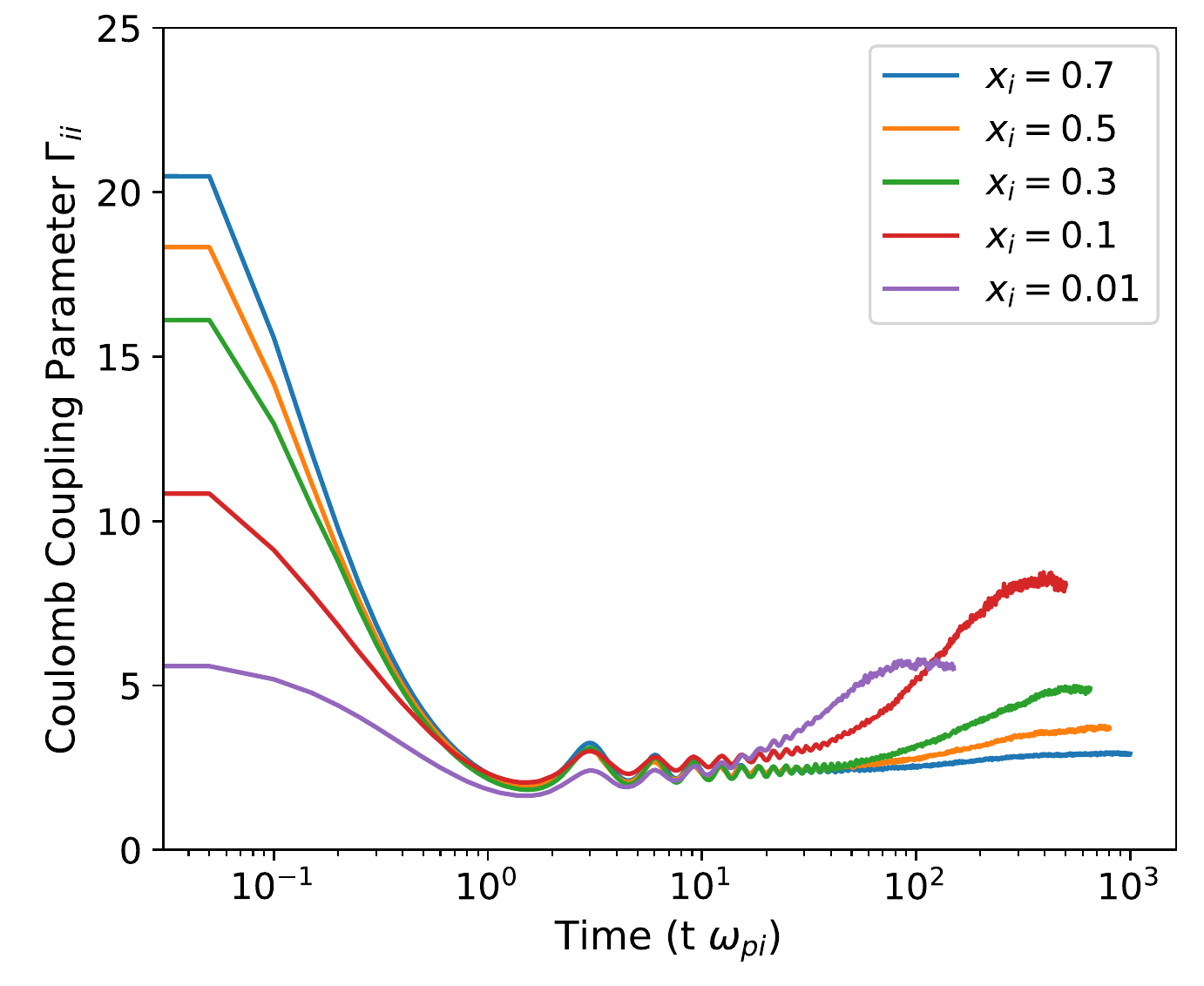}
    \caption{Evolution of the Coulomb coupling parameter $\Gamma_{ii}$ during the discharge at different ionization fractions.}
    \label{fig:gamma_t}
\end{figure}

A simple estimate for the magnitude and timescale of DIH can be obtained by considering the motion of a typical ion. 
In the limit that the initial configuration is random in space, and the final distribution is perfectly ordered (a lattice), ions will move an average distance of approximately $a_{ii}/2$. 
The electrostatic potential change in moving from a position $r_1 = a_{ii}/2$ to $r_2 = a_{ii}$ is $\Delta \phi = \phi_2 - \phi_1 = -eZ/(4\pi \epsilon_0 a_{ii})$. 
This will decrease the total potential energy of the ions by a factor of $\Delta$U $\approx -Z^2 e^2/(4 \pi \epsilon_0 a_{ii})$. Since energy is conserved during the NVE simulation, this leads to a corresponding increase in the ion kinetic energy $\Delta$K $ = -\Delta$U $ \approx Z^2 e^2/(4 \pi \epsilon_0 a_{ii})$. For ionization fractions above approximately $10^{-4}$, the initial value of $\Gamma_{ii}$ based on room temperature (when the ionization pulse is applied) is considerably larger than 1, so the change in the ion kinetic energy due to DIH is much greater than the kinetic energy before the pulse.  
Estimating the temperature after DIH as $\frac{3}{2} k_B T \approx \Delta$K, the Coulomb coupling parameter from equation~(\ref{eq:gamma}) is  $\Gamma_{ii} \approx 1.5$. 
In all cases simulated, DIH is observed to increase the ion temperature until a critical value of approximately $\Gamma_{ii} = 1.9$ is reached; see figure~\ref{fig:gamma_t}. 
This is quite close to the estimate of 1.5. 
The value of the temperature directly after DIH is the maximum value observed, $T^{\max}_i$, as ions subsequently cool due to collisional relaxation with neutrals. 

The timescale for DIH can be estimated from the time it takes an ion to move from a position $r_1 = a_{ii}/2$ to $r_2 = a_{ii}$ due to the Coulomb repulsion of a nearby ion. 
Considering Newton's equation of motion, $m d^2x/dt^2 = eE$ and approximating $d^2x/dt^2 \sim \Delta x/\Delta t^2 \sim (a_{ii}/2)/\Delta t^2$, and $eE \sim e^2/(4\pi \epsilon_o a_{ii}^2)$, the characteristic timescale for this process is the ion plasma period $\Delta t \approx \omega_{pi}^{-1}$. 
This also agrees well with the simulations, where the maximum ion temperature is reached approximately $1.5\omega_{pi}^{-1}$ after the ionization pulse; see figure~\ref{fig:gamma_t}. 



Disorder induced heating has been previously observed in ultracold neutral plasmas formed by photoionizing laser-cooled atoms~\cite{doi:10.1063/1.3366240}.
Immediately after ionization ions have a very small kinetic energy since the neutral gas they are formed from was at millikelvin temperature. 
However, they have significant excess potential energy since the ionization has changed the potential energy landscape to that associated with long-range interactions. 
As the excess potential energy is converted into kinetic energy the ions heat to a state where $\Gamma_{ii} \approx 1$ over a timescale of $1 \omega_{pi}^{-1}$, just as described above. The process observed here is essentially the same. 
One important difference is that the ionization pulse in an atmospheric pressure plasma typically only partially ionizes the gas, whereas near total ionization is common in an ultracold neutral plasma. 

An implication is that DIH is only expected to be important if the ionization fraction is high enough. 
Heating occurs only when the initial ion state (just after ionization) satisfies $\Gamma_{ii} > 1$. 
Otherwise, the kinetic energy gained by DIH is small compared to the initial kinetic energy. Furthermore, the basic mechanism isn't expected to apply since the ions remain in a weakly coupled disordered state even after ionization. 
Assuming ions are born from neutral gas at room temperature, $\Gamma_{ii} > 1$ requires that $x_i \gtrsim 10^{-4}$. 
Thus, at room temperature atmospheric pressure gas conditions, DIH is expected to occur only if the ionization fraction is larger than one part in ten thousand $x_i \gtrsim 10^{-4}$. 

\begin{figure}
    \centering
    \includegraphics[width=\linewidth]{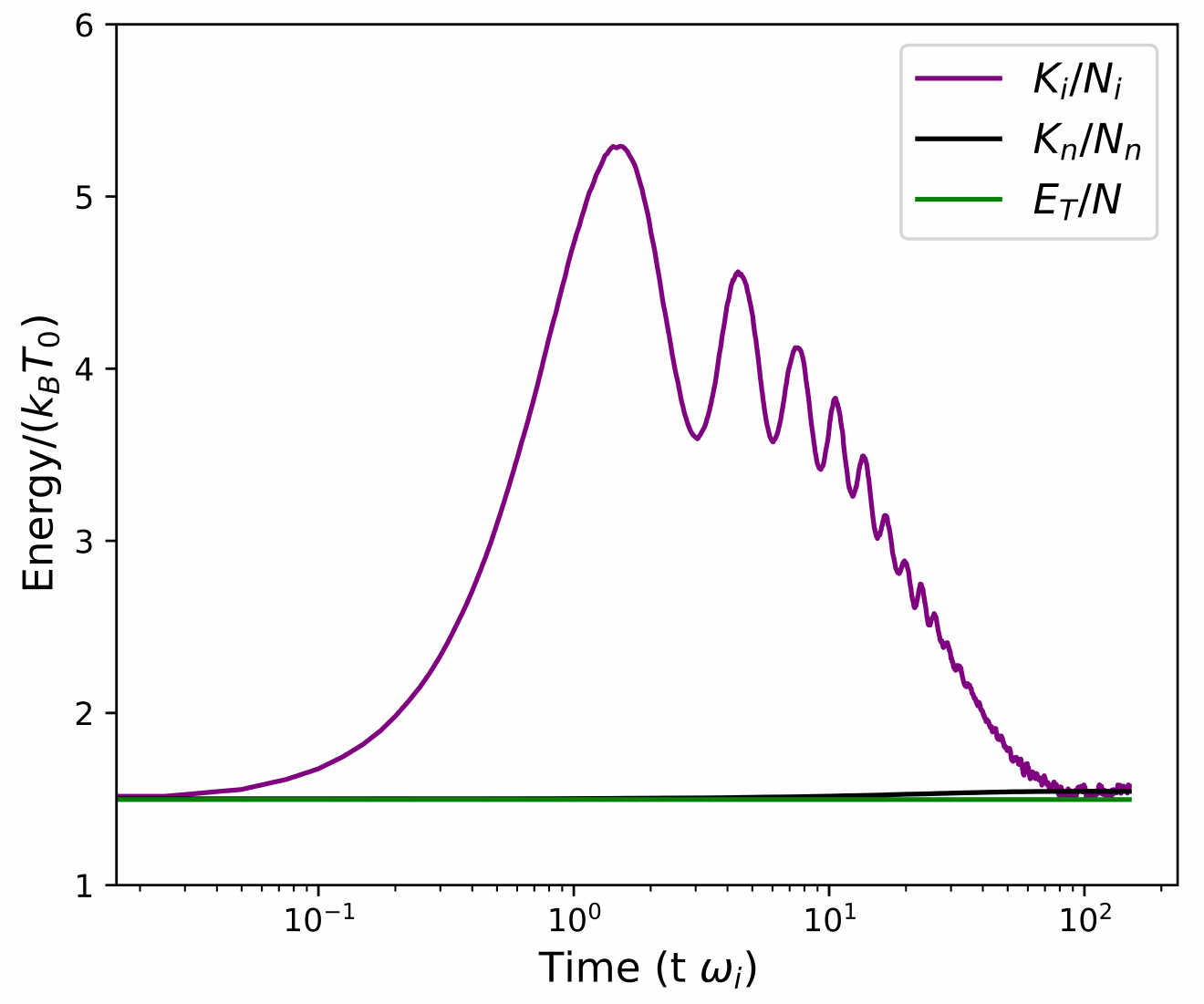}
    \caption{Ion kinetic energy ($K_i$), neutral kinetic energy ($K_n$) and total energy $E_T$ for a discharge at atmospheric pressure and an ionization fraction of $x_i=0.01$. Each energy is normalized by the number of ions ($N_i$), neutral atoms ($N_n$) and total number of particles $N$ respectively and the kinetic energy at room temperature $T_0=300 \; K$. Large fluctuations were observed in the ion kinetic energy due to the oscillations in the Coulomb potential energy after the DIH.}
    \label{fig:energy_fluctuations_xi_0_01}
\end{figure}

After DIH, ions overshoot their equilibrium positions leading to oscillations of the Coulomb potential energy near the ion plasma frequency. Since the total energy is conserved during the NVE simulation and the ions are strongly coupled after the DIH, those oscillations translate to observable kinetic energy fluctuations. 
This is shown in figure \ref{fig:energy_fluctuations_xi_0_01} from the simulation with an ionization fraction of 0.01. Such fluctuations are only noticeable in the strongly coupled regime, where the ion kinetic energy is comparable to the Coulomb potential energy ($\Gamma \approx  1$). These fluctuations were observed to damp over a period of time between 2$\omega^{-1}_p$ and 50$\omega^{-1}_p$ depending on the ionization fraction; see figure \ref{fig:Ti_t_xi}. The presence of large fluctuations in the ion temperature after DIH has also been observed in ultracold neutral plasmas experiments~\cite{doi:10.1063/1.3366240} and MD simulations~\cite{KuzminPRL2002}.

\subsection{Ion-Neutral Temperature Relaxation}

After disorder-induced heating, ion and neutral temperatures equilibrate due to ion-neutral collisions. 
This causes the ions to cool and the neutrals to heat by an amount and at a rate that depends on the ionization fraction. 
Ion cooling increases the ion-ion coupling strength, as shown in figure~\ref{fig:gamma_t}. This leads to an ion-ion coupling strength that is larger than one, $\Gamma_{ii}>1$, after the ion and neutral temperatures relax for all ionization fractions simulated. 
It is also noteworthy that fast neutral heating can be significant, especially when the ionization fraction is high. 
Neutral heating is a common observation in atmospheric pressure plasma experiments \cite{Mintoussov_2011,Popov_2016,Pai_2010,Popov_2011}.

\begin{figure}
    \centering
    \includegraphics[width=\linewidth]{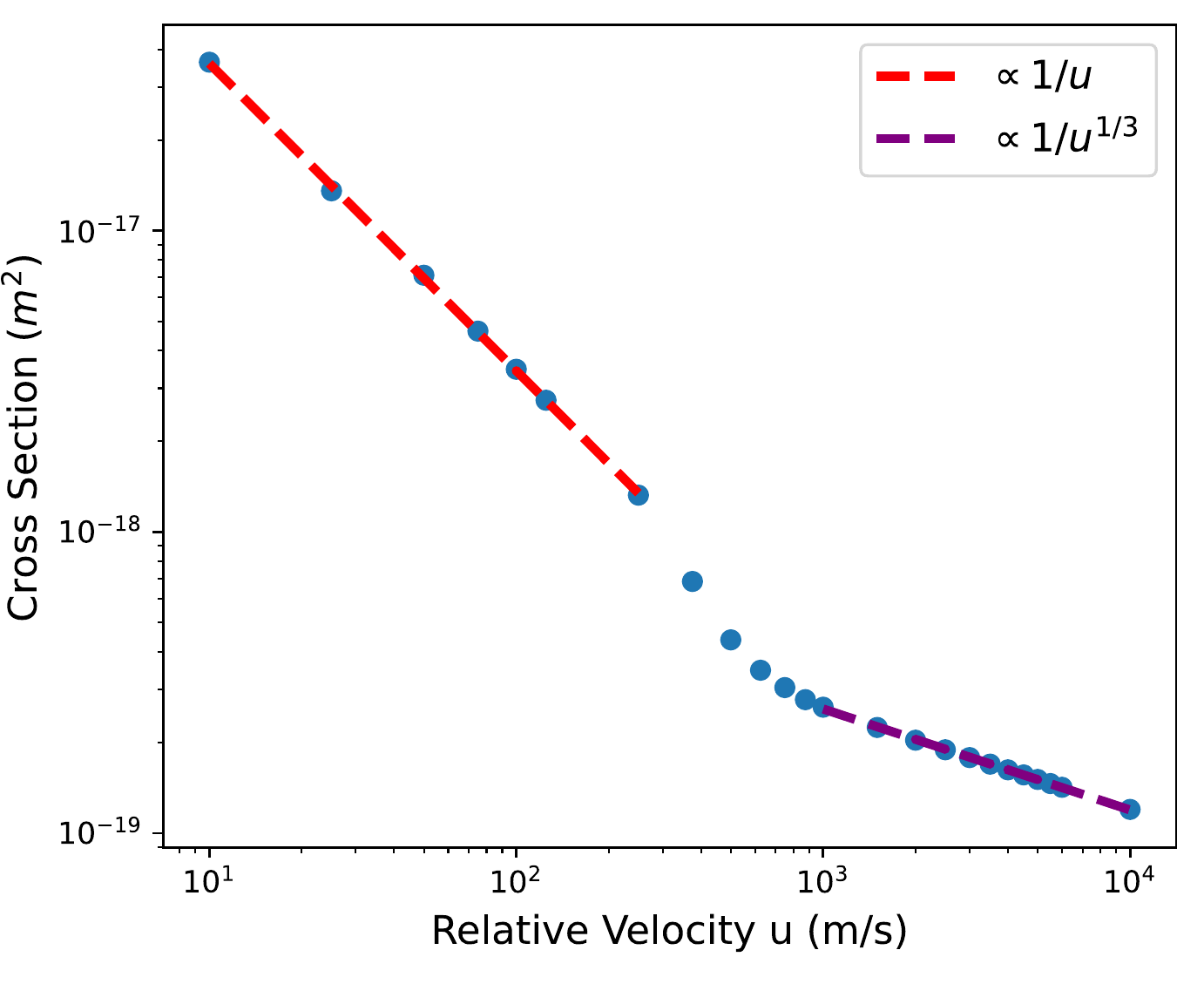}
    \caption{Ion-neutral momentum transfer cross section at different relative velocities, obtained for the charge induced dipole potential used in the simulations.}
    \label{fig:sigma_u}
\end{figure}

Ion-neutral temperature relaxation was modeled using standard methods based on the Boltzmann equation~\cite{Ferziger}. 
The cross section for ion-neutral interactions was computed based on two-body collisions interacting through the charge-induced dipole potential from equation~(\ref{eq:induced_numerical}). 
The Boltzmann-based approach is expected to be valid since the ion-neutral interaction is in a weakly coupled regime; see figure~\ref{fig:gamma_diagram}. 
The resulting temperature relaxation rate is described by
\begin{equation}
    \frac{dT_i}{dt} = -\frac{3}{2} \nu_{in}(T_i - T_n) ,
    \label{eq:Ti_eq}
\end{equation}
\begin{equation}
    \frac{dT_n}{dt} = \frac{3}{2} \nu_{ni}(T_i - T_n),
    \label{eq:Tn_eq}
\end{equation}
and
\begin{equation}
    \nu_{ss'}=\frac{4n_{s'}\bar{v}_{ss'}}{3} \int_{0}^{\infty}  dg \;\; Q^{(1)}_{ss'}(g) g^5 e^{-g^2} .
    \label{eq:nu_ss_eq}
\end{equation} 
Here, $T_i$ and $T_n$ are ion and neutral temperatures respectively, $\nu_{ss'}$ is the energy transfer collision frequency between the species $s$ and $s'$, $g=u/\bar{v}_{ss'}$ where $u$ is the relative velocity and $\bar{v}_{ss'}^2=2k_BT_s/m_s + 2k_BT_{s'}/m_{s^\prime}$, where $m_s$ and $m_{s^\prime}$ are both the Ar mass, and $Q^{(1)}_{ss'}$ is the momentum transfer cross section 
\begin{equation}
    Q^{(1)}_{ss'} = 2\pi \int^\infty_0 [1-\cos(\chi)]bdb, 
    \label{eq:cross_section}
\end{equation}
where
\begin{equation}
    \chi = \pi - 2b \int^\infty_{r_0} \frac{dr/r^2}{\sqrt{1 - \frac{b^2}{r^2} - \frac{2\phi_{in}(r)}{m_{ss'}u^2}}} 
    \label{eq:scattering_angle}
\end{equation}
is the scattering angle. 
Here, $b$ is the impact parameter, $r_0$ is the distance of closest approach obtained from the largest root of the denominator in equation~(\ref{eq:scattering_angle}), $r$ is the radial distance, $m_{ss'}$ is the reduced mass, $u$ is the relative velocity and $\phi_{in}$ is the charge induced dipole potential from equation~(\ref{eq:induced_numerical}) with $r_{\phi}=0.133a_{in}$ computed from equation~(\ref{eq:cross_section}). 
The ion-neutral momentum transfer cross section obtained is shown in figure \ref{fig:sigma_u}. 

\begin{figure}
    \centering
    \includegraphics[width=\linewidth]{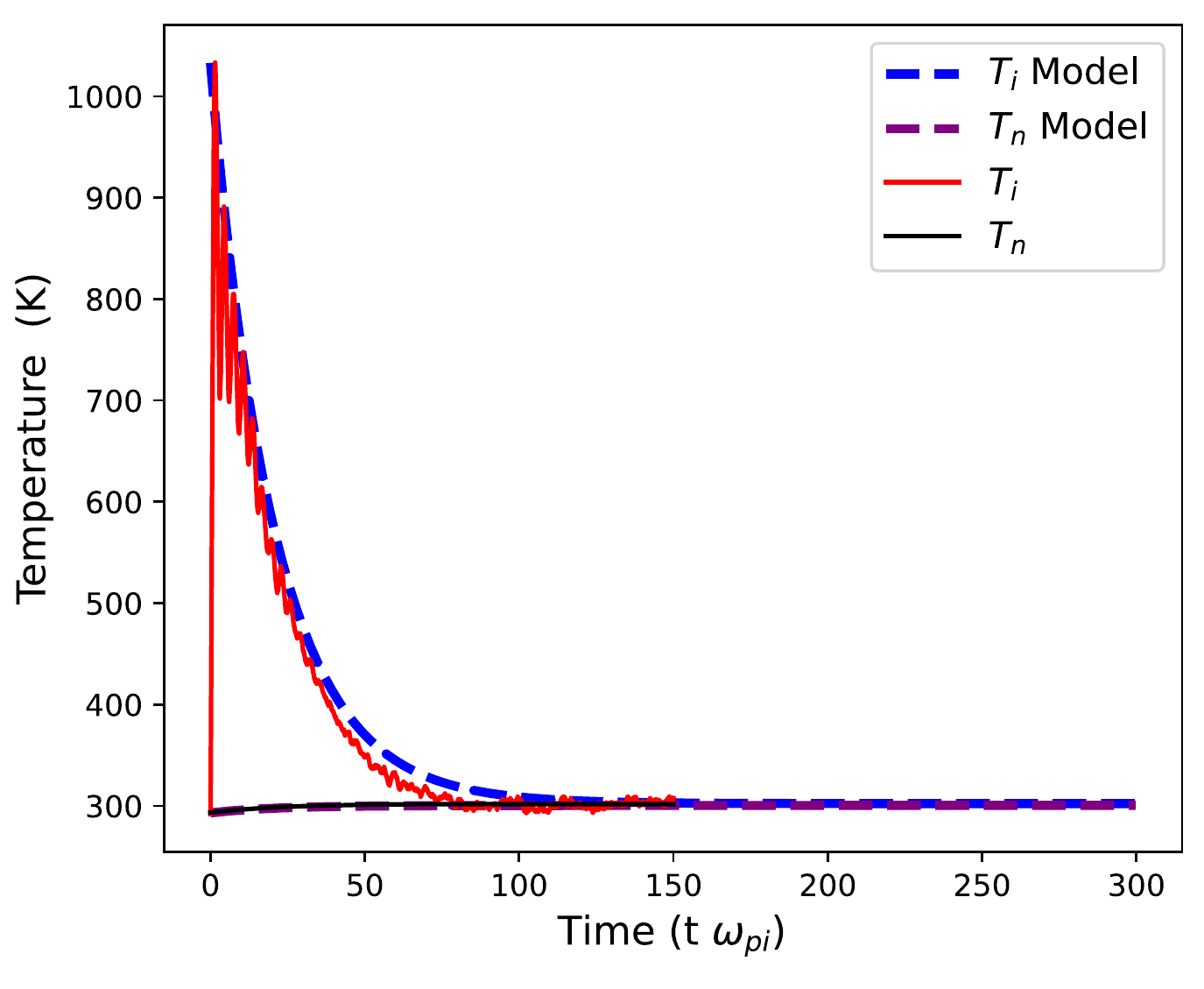}
    \caption{Evolution of the ion temperature for a discharge with $x_i=0.01$ along with the model described in equations~(\ref{eq:Ti_eq})--(\ref{eq:nu_ss_eq}).}
    \label{fig:Ti_t_theory}
\end{figure}

The evolution of the ion temperature obtained using equations~(\ref{eq:Ti_eq})--(\ref{eq:nu_ss_eq}) is shown in figure \ref{fig:Ti_t_theory} for the discharge at an ionization fraction of 0.01. 
Here, the initial temperature used in the model was taken from the value of the MD simulations after DIH. 
The predictions show good agreement with the MD simulations in both the equilibrium temperature as well as the relaxation time. 
At other ionization fractions, the equilibrium temperature increases with ionization fraction due to smaller neutral atom densities and due to the larger ion temperature achieved due to the DIH. The equilibrium temperature was higher than room temperature and the corresponding coupling parameter was smaller compared to the the initial value of $\Gamma_{ii}$ with the exception of the smallest ionization fraction of 0.01, where the large neutral density led to an equilibrium temperature similar to room temperature; see figure~\ref{fig:Ti_t_xi}.

\subsection{Three-Body Recombination}

\begin{figure}
    \centering
    \includegraphics[width=\linewidth]{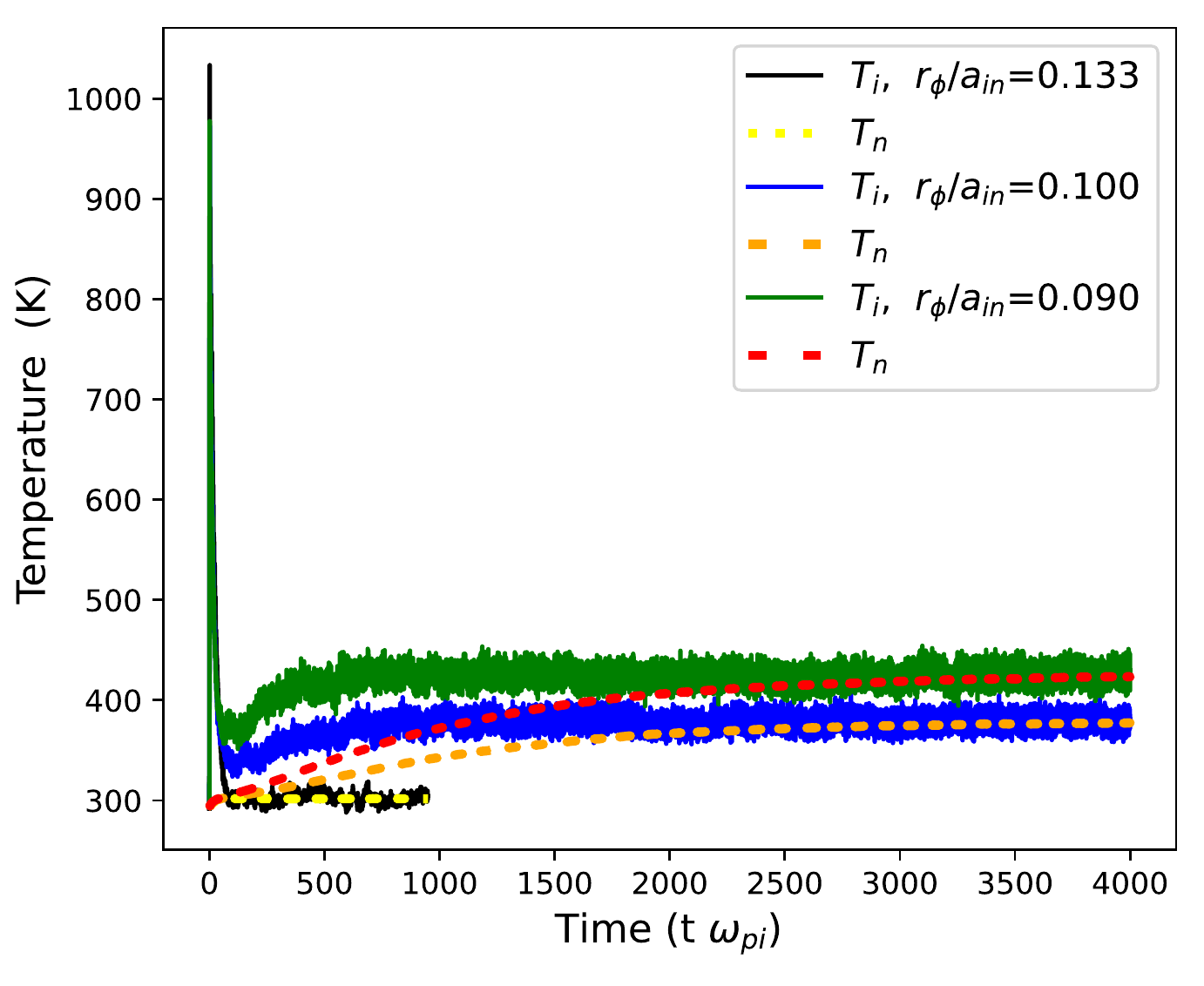}
    \caption{Evolution of the temperature for a discharge with a ionization fraction $x_i=0.01$ for different $r_{\phi}$ values. }
    \label{fig:x_0_01_recombination_comp_linear}
\end{figure}

\begin{figure}
    \centering
    \includegraphics[width=\linewidth]{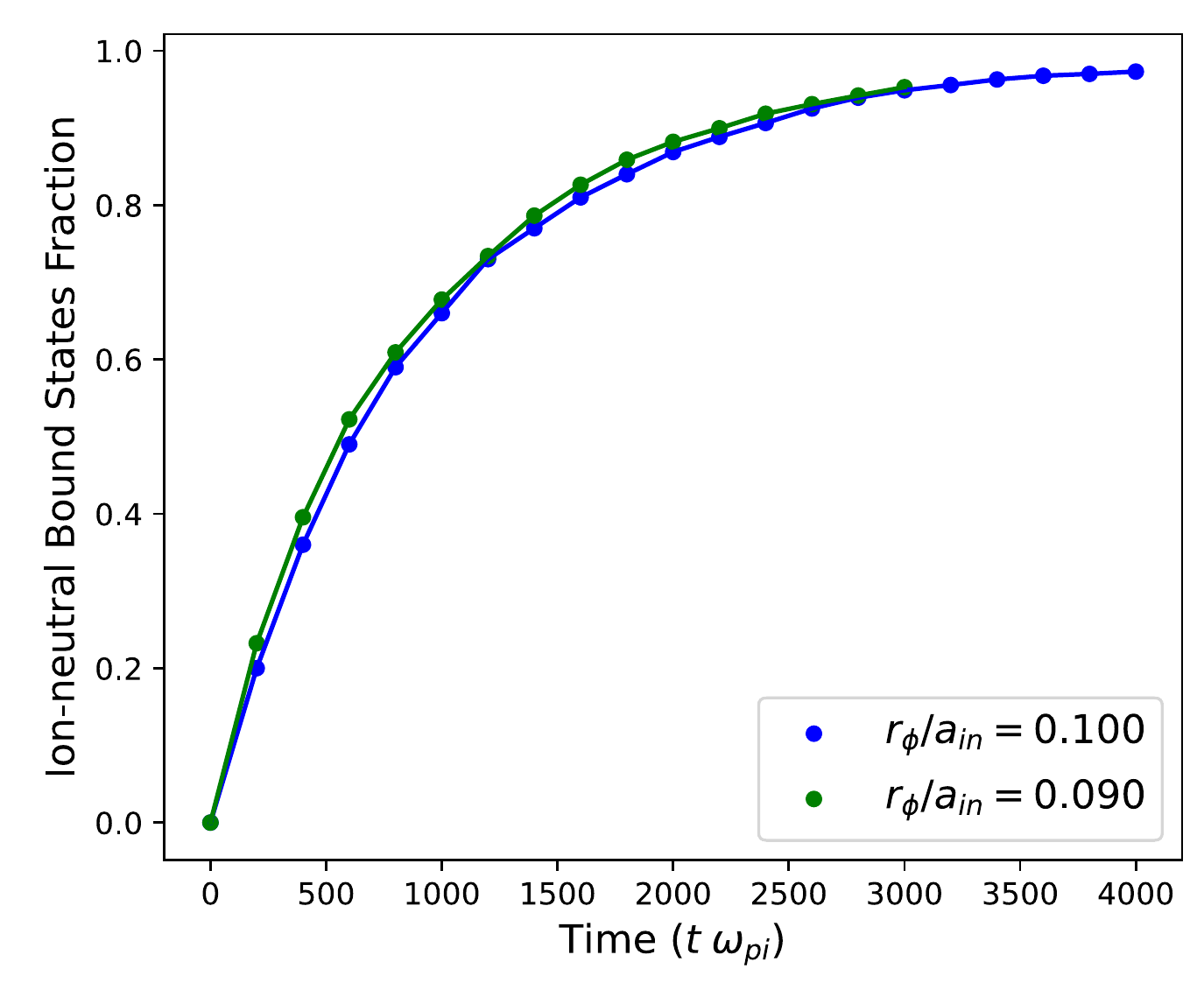}
    \caption{Ion-neutral three-body recombination rate for a discharge with a ionization fraction $x_i=0.01$ for different $r_{\phi}$ values. }
    \label{fig:x_0_01_recombination_fraction}
\end{figure}

As described in section \ref{sec:MD_simulations}, $r_{\phi}$ was used as a numerical convergence parameter in the non-equilibrium simulations and the temperature evolution over the timescale of DIH and ion-neutral equilibration was independent of $r_\phi$. 
However, as shown in the Appendix \ref{sec:A}, for  $r_{\phi}/a_{in}<0.133$ neutral particles start to orbit ions, which forms bound states that can cause heating over a much longer timescale than the DIH or ion-neutral relaxation. In order to compare these time scales, we repeated the non-equilibrium simulation on a longer time scale with different values of $r_{\phi}/a_{in}$. Simulations at an ionization fraction of $x_i=0.01$ and $r_{\phi}/a_{in}$ values of $0.100$ and $0.090$ were carried out with a total simulation time of $t\times \omega_{pi} = 4000$ and compared with a simulation at the same ionization fraction but using $r_{\phi}/a_{in}=0.133$.
As shown in figure \ref{fig:x_0_01_recombination_comp_linear}, the simulations show a similar ion temperature evolution immediately after the ionization pulse, including DIH followed by ion-neutral temperature relaxation through collisions. However, at $t\times \omega_{pi} \approx 50$, bound states start to form when $r_\phi/a_{in} < 0.133$, which increases the ion temperature when compared to the same simulation with $r_{\phi}/a_{in}=0.133$. 
This is due to particles becoming trapped in the potential well of the charge-induced dipole potential, which exchanges a reduction in the potential energy with an increase in the kinetic energy. 
The increase of the fraction of ion-neutral bound states over time is shown in figure \ref{fig:x_0_01_recombination_fraction} for both $r_{\phi}/a_{in}$ values. 
The recombination fraction was determined from a histogram of nearest neighbor positions, as described in appendix~\ref{sec:A}. 
The additional heating due to the ion-neutral three-body recombination causes the temperature to increase by approximately 75 K and 125 K, reaching a value near 400 K and 425 K at $t \times \omega_{pi} = 4000$ for $r_{\phi}/a_{in}$ values of $0.100$ and $0.090$ respectively. 
This corresponds to a time at which $97\%$ of ions have at least one neutral atom orbiting them. It was observed that the timescale of the ion-neutral three-body recombination is much longer than ion-neutral temperature relaxation; $\sim 1000 \omega_{pi}^{-1}$ corresponds to $\sim 100$s ns.

\subsection{Radial Distribution Function}

The radial distribution gives information of how the density varies with the distance from a reference particle and it can be used to look for signatures of strong correlations. 
It also determines the non-ideal components of thermodynamic parameters \cite{HANSEN201313}. 
Figure~\ref{fig:gr_0_01} shows that the radial distribution for the interactions between ions corresponds to that of the one-component plasma (OCP) at the same equilibrium coupling parameter. 
Here, the ionization fraction was 0.01. 
This result shows that the ion-ion interactions are in a strongly coupled regime and the radial distribution function shows signatures of these strong correlations. Since the radial distribution function corresponds to an OCP at the same $\Gamma_{ii}$ the ion-ion interactions are not screened by the presence of neutral atoms. This behaviour was observed for all the ionization fractions simulated. 
The ion-ion radial distribution function also exhibits a Coulomb hole and correlation peak, which are common features of strongly coupled plasmas~\cite{OttPOP2014}.

\section{Model for Maximum and Equilibrium Ion Temperature.}\label{sec:model}

\begin{figure}
    \centering
    \includegraphics[width=\linewidth]{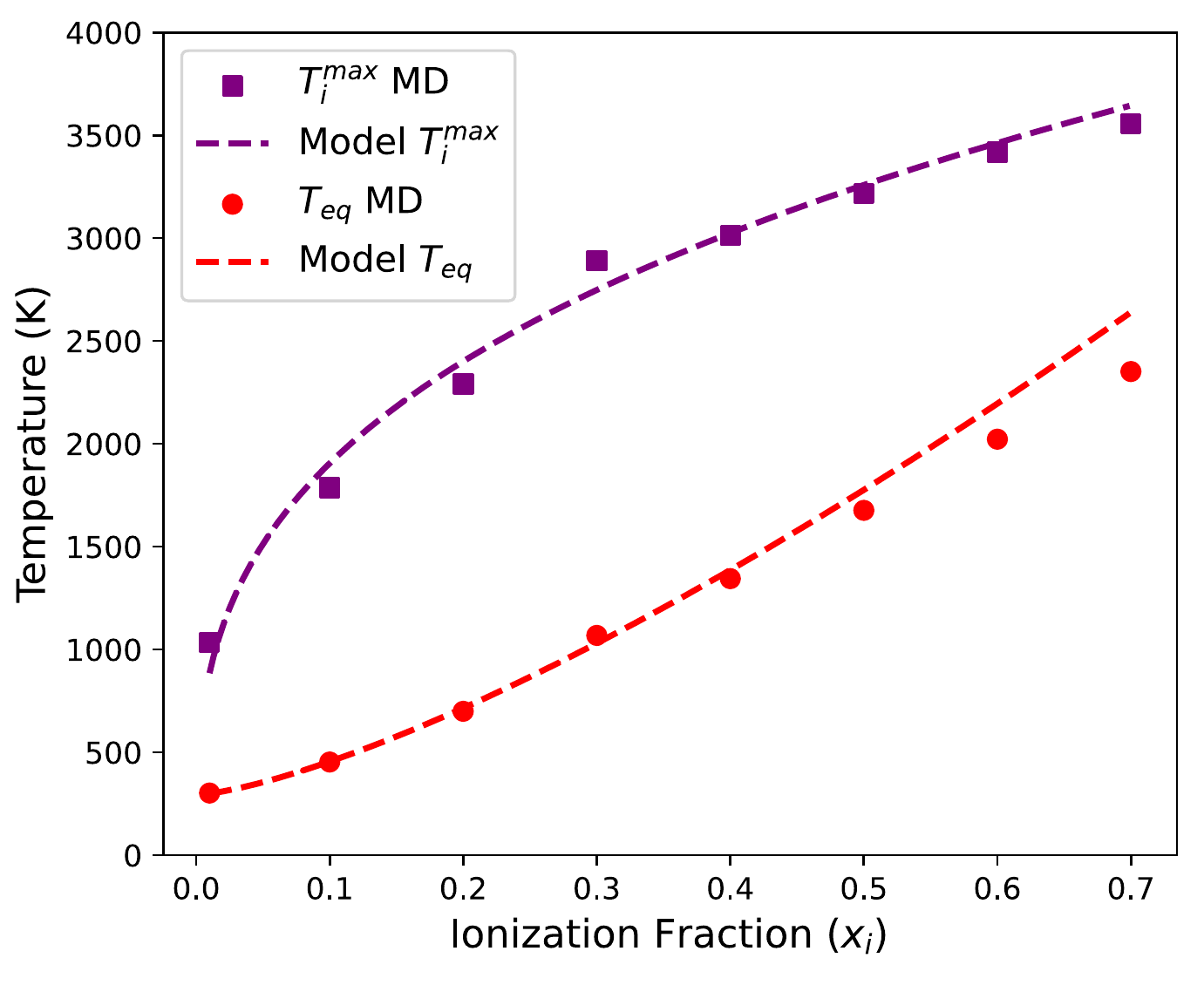}
    \caption{Variation of the maximum ion temperature and equilibrium temperature with the ionization fraction from the MD simulations and the model.}
    \label{fig:T_xi_theory}
\end{figure}

The maximum and equilibrium ion temperatures can be estimated using simple energy conservation arguments. Since the increase in the ion temperature is due to DIH, where the ions travel a distance $\approx\;a_{ii}$ over a plasma period $\omega^{-1}_p$, the maximum ion temperature corresponds to that which makes the ion-ion coupling parameter approximately unity. 
Utilizing the simulation result that $\Gamma_{ii} \approx 1.91$ at the peak temperature and equation~(\ref{eq:gamma}), the maximum ion temperature is estimated to be 
\begin{equation}
    T^{\max}_i = \frac{1}{1.91} \frac{Z^2e^2}{4\pi\epsilon_0k_B} \frac{1}{a_{ii}}
    \label{eq:Ti_max}
\end{equation} where $a_{ii}=(3/4\pi x_i n)^{1/3}$ is the average interparticle spacing between ions. 
The maximum ion temperature estimated with the equation~(\ref{eq:Ti_max}) shows good agreement with the results obtained from the MD simulations, as shown in figure~\ref{fig:T_xi_theory}.

\begin{figure}
    \centering
    \includegraphics[width=\linewidth]{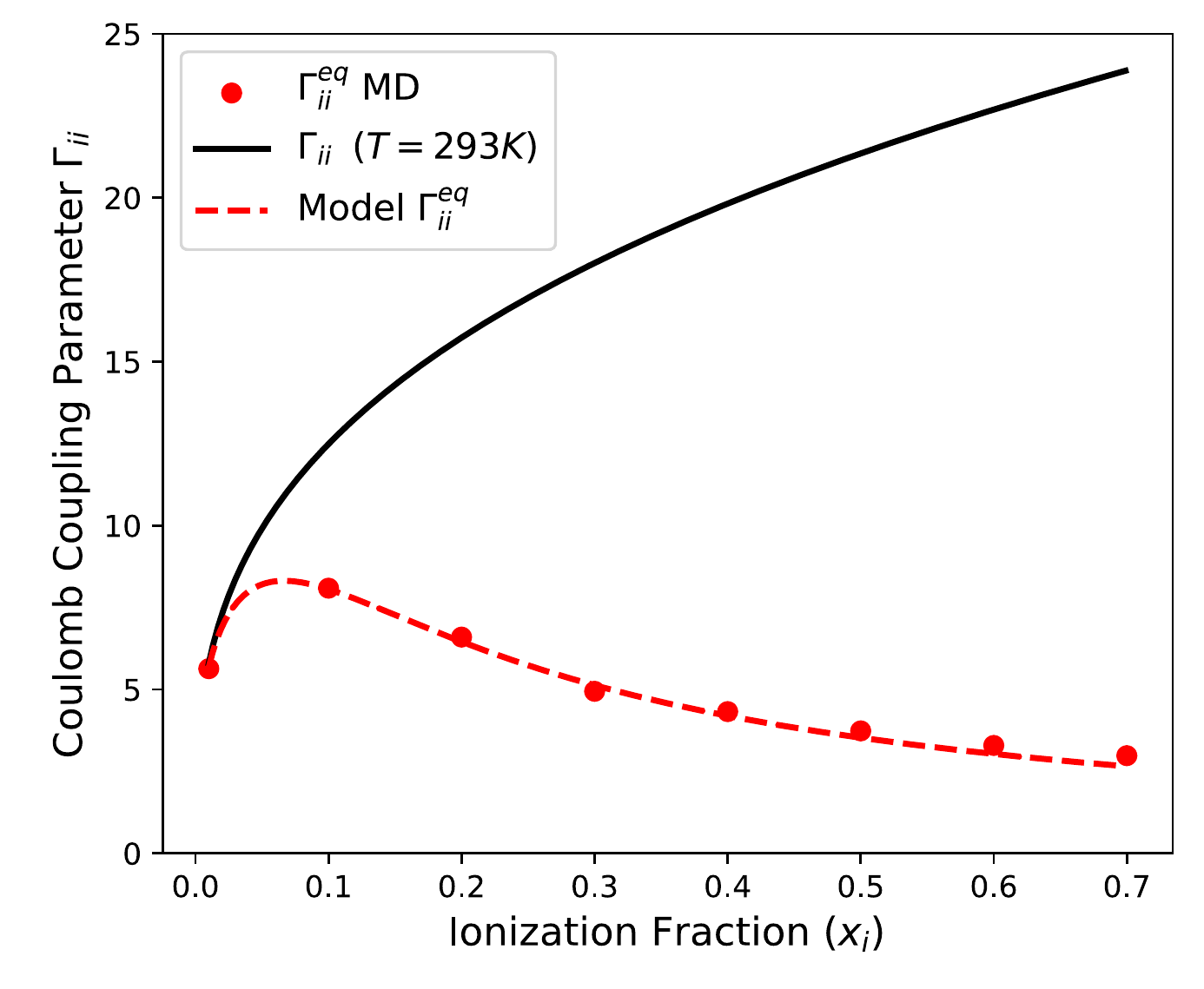}
    \caption{The Coulomb coupling parameter at equilibrium is smaller than the coupling parameter at room temperature without accounting for DIH. At small ionization fractions $\Gamma^\textrm{eq}_{ii}(T^\textrm{eq},x_i)$ converges to $\Gamma_{ii}(T_0,x_i)$ where $T_0$ is the room temperature.}
    \label{fig:gamma_xi_theory}
\end{figure}

If we assume that the total kinetic energy is conserved after DIH until thermodynamic equilibrium is reached, it is possible to write a simple energy equation to obtain the equilibrium temperature $T_\textrm{eq}$, i.e., the temperature after ion-neutral thermal equilibration, 
\begin{equation}
    T_\textrm{eq} = x_i \; T^{\max}_i + (1-x_i) \; T_n(t=0)
    \label{eq:T_eq}
\end{equation} where $T_n$ is the neutral atom temperature at the beginning of the simulation, $x_i$ is the ionization fraction and $T^{\max}_i$ is the maximum ion temperature obtained with equation~(\ref{eq:Ti_max}). As shown in figure \ref{fig:T_xi_theory}, the values of $T_\textrm{eq}$ and $T^{\max}_i$ obtained using the model show good agreement with MD simulations over a broad range of ionization fractions. Furthermore, the expected coupling parameter with and without accounting for the DIH can be calculated by using the correct equilibrium temperature and the room temperature respectively, as shown in figure \ref{fig:gamma_xi_theory}. The values predicted for $\Gamma^\textrm{eq}_{ii}$ show good agreement with the results from the MD simulations. It is noticeable how the coupling parameter at equilibrium is smaller than what would be predicted using room temperature, since the DIH increases the equilibrium temperature of the system at large ionization fractions. However, at small ionization fractions the values for $\Gamma_{ii}(T_0,x_i)$ and $\Gamma^\textrm{eq}_{ii}(T_\textrm{eq},x_i)$ match due to the large neutral density compared to the ion density.  

\section{Comparison with Experiment\label{sec:comparison}}

Previous measurements have observed considerable neutral gas heating in ns pulsed spark discharges at atmospheric pressure. 
For example van der Horst \emph{et al} measured a neutral gas temperature of 750 K at at time 1 $\mu$s after ignition in a N$_2$/H$_2$O mixture~\cite{vanderHorst2012}.
They also showed that the pressure due to the discharge increased to 3 bar and the corresponding ionization fraction was $x_i=0.16$. Using the model for the equilibrium temperature from equation~(\ref{eq:T_eq}) and the pressure and ionization fraction measured in \cite{vanderHorst2012}, we calculate an expected ion-neutral equilibrium temperature of 765 K with a relaxation time of 2.6 ns. 
This agrees quite closely with the experimental measurement. 
The original reference suggested that the gas heating may be due to elastic collisions between electrons and $N_2$. 
However, for the conditions of this experiment, the energy transfer time due to elastic collisions is about 0.5~$\mu$s. 
This is quite a slow process where the relaxation time is near the time at which the temperature was measured. 

The heating mechanism proposed here occurs over the first 13 ps after ignition for the pressure and ionization fraction measured in \cite{vanderHorst2012}. The subsequent ion-neutral temperature relaxation occurs over 2.6 ns after the ignition. The noticeable separation of timescales shows that a better time resolution for experiments could help to identify the responsible heating mechanisms.  

The results presented here suggest that the DIH and subsequent ion-neutral relaxation should be considered as a possible mechanism for the observed neutral heating.



\section{Conclusions}

This work demonstrates that ions are strongly coupled in atmospheric pressure plasmas and that strong correlations influence ion and neutral temperature evolution. 
First-principles MD simulations reveal that the ion and neutral temperatures in CAPPs are influenced by an associated disorder induced heating process. Disorder-induced heating is not important in weakly coupled plasmas, but can be a dominant effect in strongly coupled plasmas. 
It causes ions to rapidly heat to temperatures that can reach several times the background neutral gas temperature. After DIH, ions thermally equilibrate with neutrals causing them to cool and the neutrals to heat by an amount and at a rate that depends on the ionization fraction. The cooling causes ions to return to a more strongly coupled state. Due to DIH and ion-neutral temperature relaxation, the final neutral gas temperature depends on the ionization degree achieved at plasma formation. We show that at atmospheric pressure, DIH is significant for ionization fractions larger than $x_i \gtrsim 10^{-4}$ and thus, it could be important for experiments that involve spark discharges, laser-produced plasmas, or other sources that achieve high ionization fractions. In parallel with these comparatively fast processes, a weaker heating of both ions and neutrals was observed over a much longer timescale due to ion-neutral three-body recombination. 
A model is presented to describe the main features of the ion and neutral temperature evolution, showing a good agreement with the neutral gas temperature measured in an experiment at atmospheric pressure \cite{vanderHorst2012}. Therefore, we show that DIH is an additional heating mechanism that can occur in CAPP applications and must be accounted for. 

One implication of this work is that modeling techniques for CAPPs may need to be reassessed. 
Current methods are largely based on solutions of a Boltzmann equation, such as PIC, or moments of the Boltzmann equation, such as multi-fluid models. 
These implicitly assume that the plasma is weakly coupled, since the Boltzmann equation only treats weakly coupled gases and plasmas. 
This may be limiting in the context of CAPPs, because fundamentally different physical behaviors arise in strongly coupled fluids. 
For instance, DIH could not be obtained by solving the Boltzmann equation, as in PIC simulations.  
Using such models may lead to a significant underestimation of the ion and neutral temperatures. 
This motivates the need for further development of theory and computational models that can account for strong coupling effects in cold atmospheric pressure plasmas.

\section{Acknowledgements}

This work was supported in part by the US Department of Energy under award no.~DE-SC0022201, and in part by Sandia National Laboratories. 
Sandia National Laboratories is a multi mission laboratory managed and operated by National Technology and Engineering Solutions of Sandia, LLC., a wholly owned subsidiary of Honeywell International, Inc., for the U.S. Department of Energy’s National Nuclear Security Administration under Contract No. DE-NA0003525. This article describes objective technical results and analysis. Any subjective views or opinions that might be expressed in the paper do not necessarily represent the views of the U.S. Department of Energy or the United States Government.


\appendix

\section{Equilibrium MD Simulations.}\label{sec:A}

\begin{figure}
    \centering
    \includegraphics[width=\linewidth]{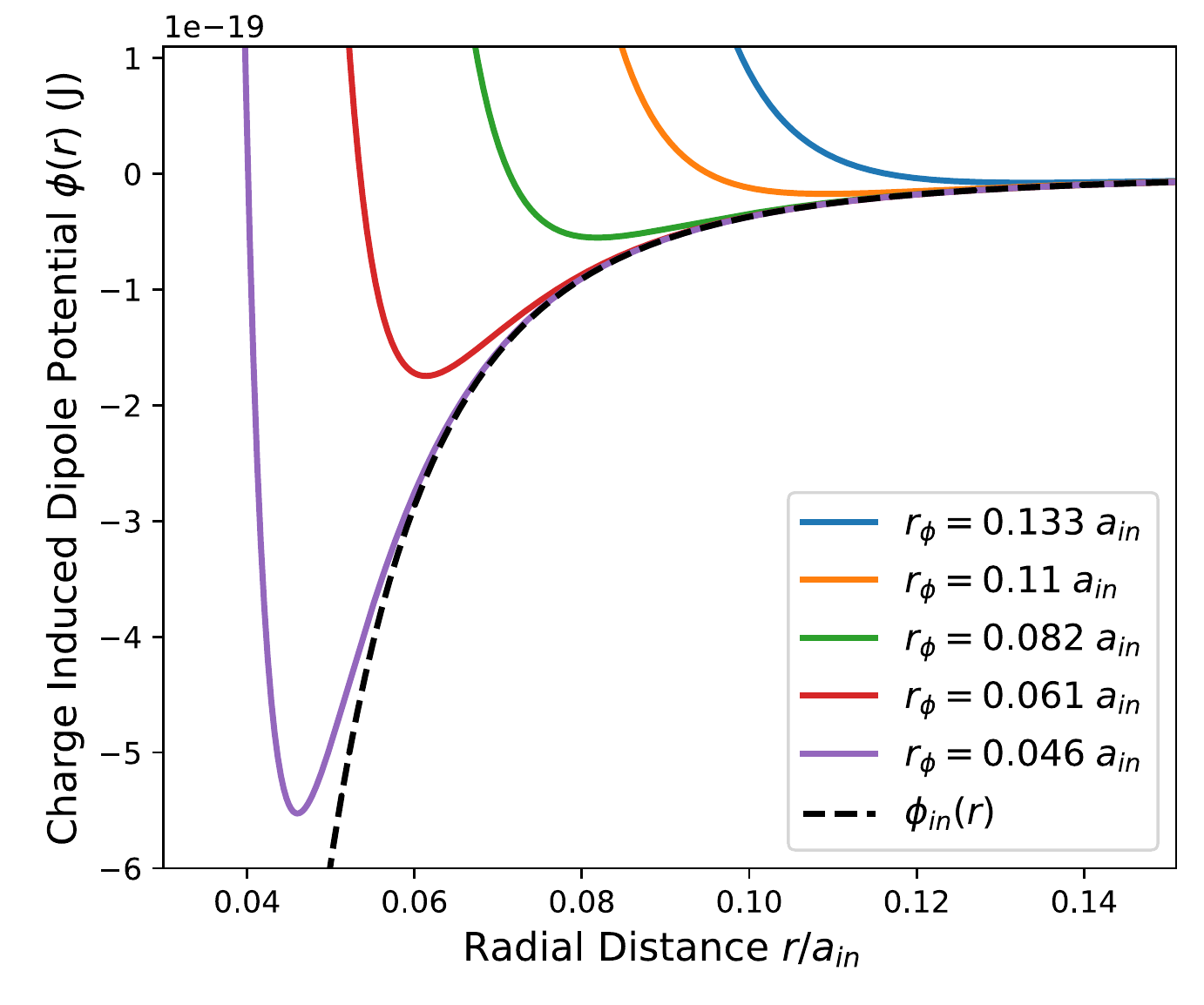}
    \caption{Numerical charge induced dipole potential for different values of $r_{\phi}$ compared to the charge induced dipole potential.}
    \label{fig:pot_in_nn}
\end{figure}

An analysis was realized varying the radius $r_{\phi}$ of the charge induced dipole potential. For this analysis the simulation setup consisted of an NVT (canonical) simulation using the Nosé–Hoover thermostat, where the temperature was set to 293 K, over 1500 plasma periods. 
This was followed by an NVE (microcanonical) simulation for 1500 plasma periods. The timestep used varied with $r_{\phi}$ since smaller $r_{\phi}$ led to larger attractive forces and smaller timesteps were required. The timesteps used varied from $5\times10^{-4}\omega_p^{-1}$ to $1\times 10^{-4}\omega_p^{-1}$ where $\omega_p^{-1}$ is the ion plasma period. The total number of particles was 10000 for all the simulations. Since the numerical charge induced dipole potential has both a repulsive and an attractive part, it presented a potential well, as shown in the figure \ref{fig:pot_in_nn}. 
Different radius and, therefore, minimum values of the charge induced dipole potential were explored. 
Once the equilibrium was reached in the NVE simulation, the distribution of minimum distances between ion-neutral and neutral-neutral pairs of particles was computed in order to study the presence of bound states. These bound states consisted of ion-neutral molecules with a spatial scale characterized by $r_{\phi}$ and $r_{LJ} \approx 2^{1/6} \;\sigma$ which corresponds to the minimum of the charge induced dipole and the Lennard Jones potentials respectively.

\begin{figure}
    \centering
    \includegraphics[width=\linewidth]{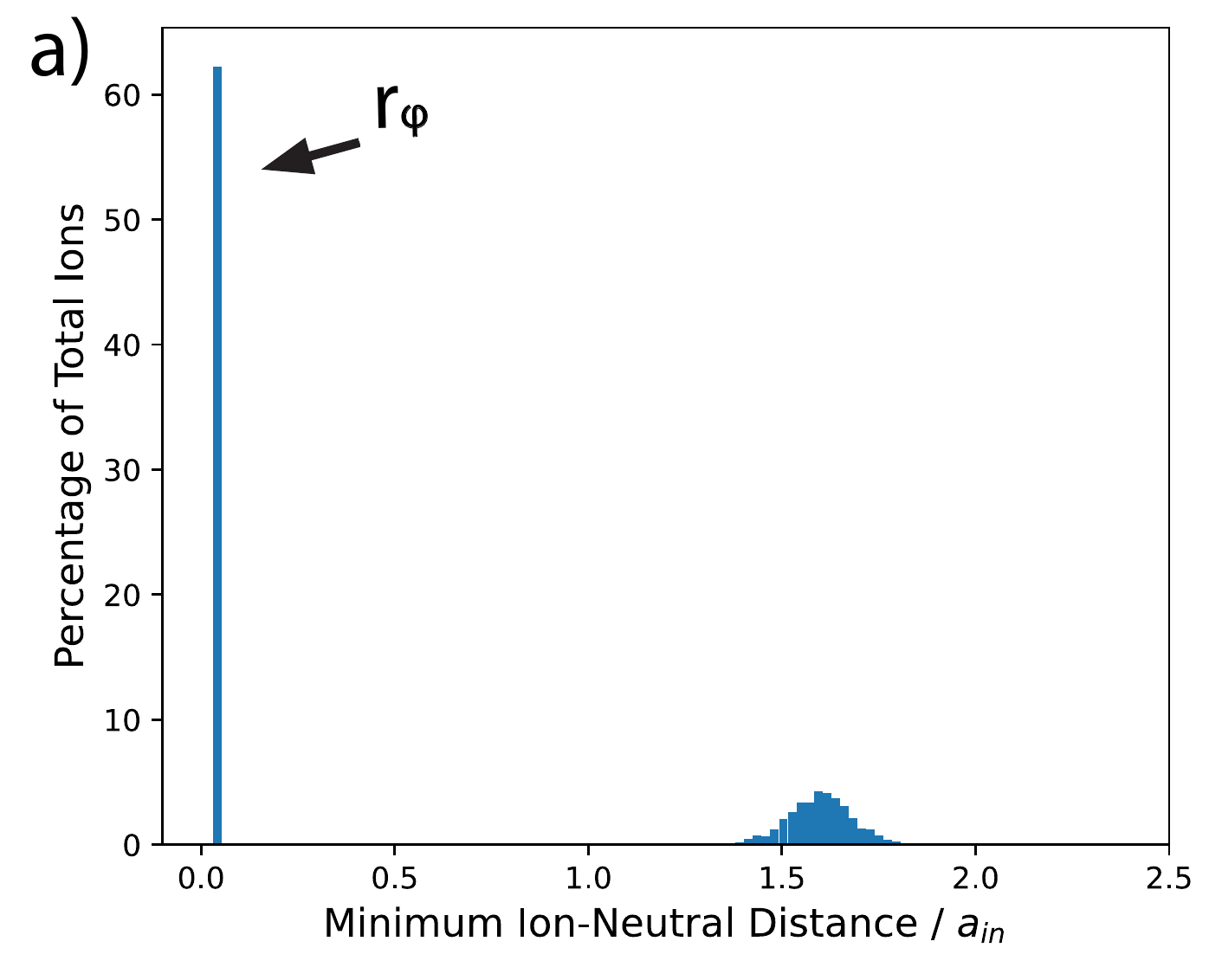}
    \includegraphics[width=\linewidth]{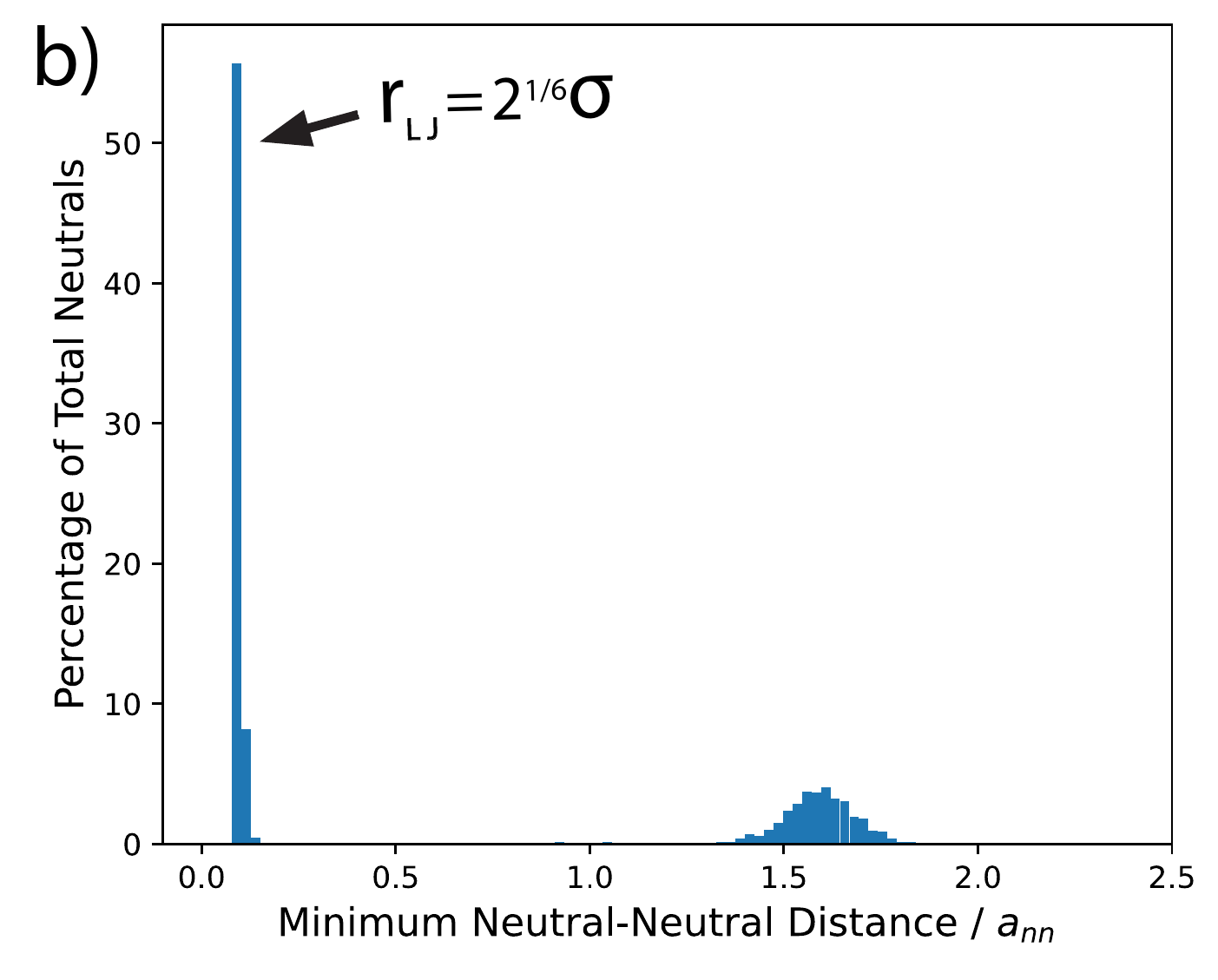}
    \caption{a) Ion-neutral and b) neutral-neutral minimum distance distribution for $r_{\phi}=0.046a_\textrm{in}$ at an ionization fraction of 0.5.}
    \label{fig:distance_distributions}
\end{figure}

It was observed that for relatively small values of $r_{\phi}$ the formation of bound states was more prominent. As it is shown in the distributions of minimum distances between ion-neutral and neutral-neutral pairs of particles after 1500 plasma periods in figure \ref{fig:distance_distributions}, for $r_{\phi}=0.046a_{in}$ and a ionization fraction of 0.5, large peaks were observed corresponding to the distances $r_{\phi}$ and $r_{LJ}$.

\begin{figure}
    \centering
    \includegraphics[width=\linewidth]{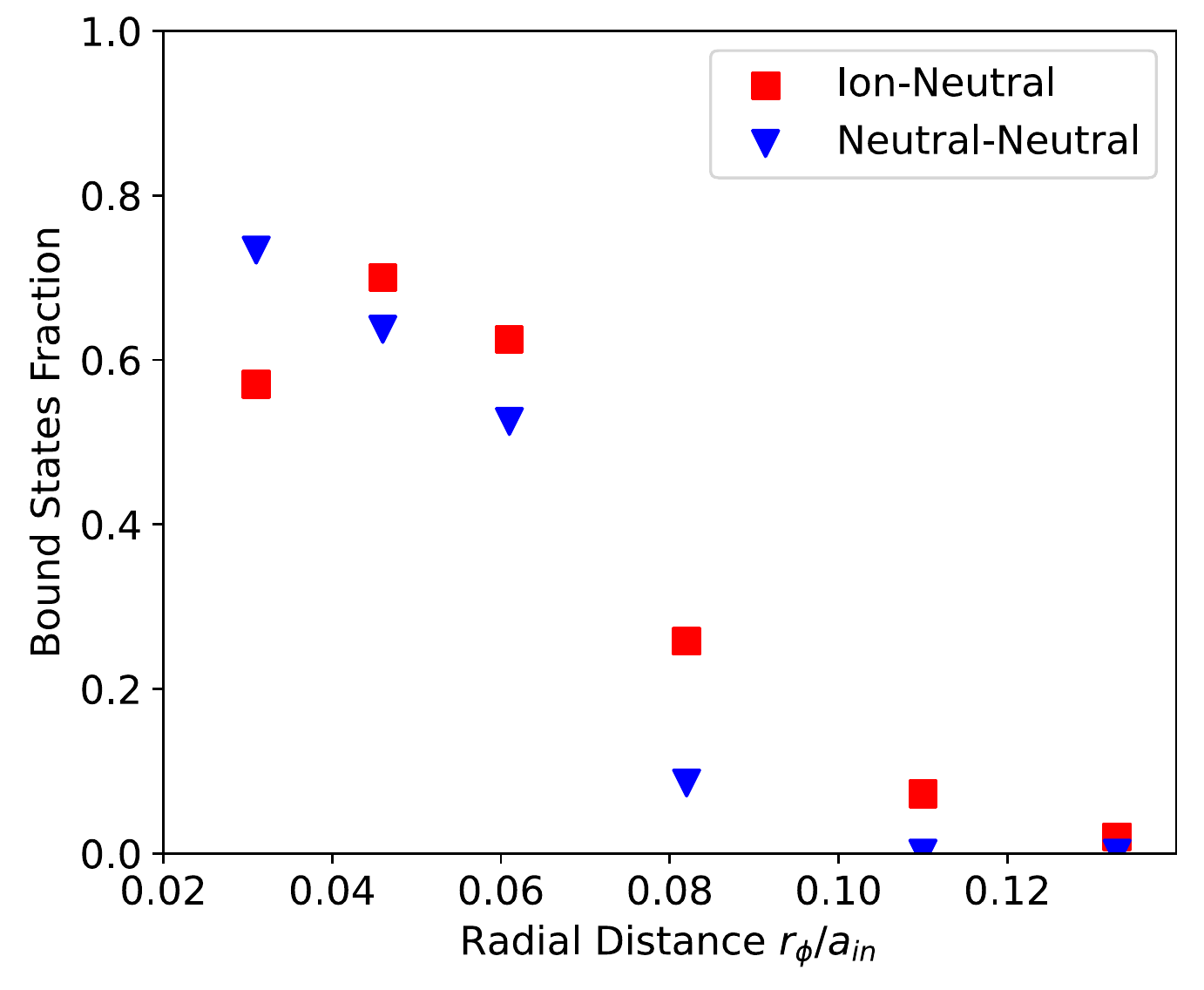}
    \caption{Fraction of ion-neutral and neutral-neutral bound states for different $r_{\phi}$ values.}
    \label{fig:bound_states}
\end{figure}

As the radius of the charge induced dipole potential was increased, the population of ion-neutral and neutral-neutral bound states decreased until it was negligible for $r_{\phi}\approx0.133a_{in}$ as shown in figure \ref{fig:bound_states}. The observed behaviour suggests that in order to avoid the formation of bound states an $r_{\phi}\approx0.133a$ must be used.

\clearpage

\bibliography{main.bbl}

\begin{thebibliography}{10}

\bibitem{app11114809}
Mária Domonkos, Petra Tichá, Jan Trejbal, and Pavel Demo.
\newblock Applications of cold atmospheric pressure plasma technology in
  medicine, agriculture and food industry.
\newblock {\em Applied Sciences}, 11(11), 2021.

\bibitem{https://doi.org/10.1002/ppap.201700085}
Patrick~J. Cullen, James Lalor, Laurence Scally, Daniela Boehm, Vladimir
  Milosavljević, Paula Bourke, and Kevin Keener.
\newblock Translation of plasma technology from the lab to the food industry.
\newblock {\em Plasma Processes and Polymers}, 15(2):1700085, 2018.

\bibitem{misra_cold_2016}
N.~N. Misra, Oliver Schlüter, and P.~J. Cullen, editors.
\newblock {\em Cold plasma in food and agriculture: fundamentals and
  applications}.
\newblock Elsevier/AP, Academic Press is an imprint of Elsevier, Amsterdam ;
  Boston, 2016.
\newblock OCLC: ocn934606423.

\bibitem{Adamovich_2022}
I~Adamovich, S~Agarwal, E~Ahedo, L~L Alves, S~Baalrud, N~Babaeva, A~Bogaerts,
  A~Bourdon, P~J Bruggeman, C~Canal, E~H Choi, S~Coulombe, Z~Donk{\'{o}}, D~B
  Graves, S~Hamaguchi, D~Hegemann, M~Hori, H-H Kim, G~M~W Kroesen, M~J Kushner,
  A~Laricchiuta, X~Li, T~E Magin, S~Mededovic Thagard, V~Miller, A~B Murphy,
  G~S Oehrlein, N~Puac, R~M Sankaran, S~Samukawa, M~Shiratani, M~{\v{S}}imek,
  N~Tarasenko, K~Terashima, E~Thomas Jr, J~Trieschmann, S~Tsikata, M~M Turner,
  I~J van~der Walt, M~C~M van~de Sanden, and T~von Woedtke.
\newblock The 2022 plasma roadmap: low temperature plasma science and
  technology.
\newblock {\em Journal of Physics D: Applied Physics}, 55(37):373001, jul 2022.

\bibitem{Neyts_2014}
E~C Neyts and A~Bogaerts.
\newblock Understanding plasma catalysis through modelling and
  simulation{\textemdash}a review.
\newblock {\em Journal of Physics D: Applied Physics}, 47(22):224010, may 2014.

\bibitem{Bogaerts_2020}
Annemie Bogaerts, Xin Tu, J~Christopher Whitehead, Gabriele Centi, Leon
  Lefferts, Olivier Guaitella, Federico Azzolina-Jury, Hyun-Ha Kim, Anthony~B
  Murphy, William~F Schneider, Tomohiro Nozaki, Jason~C Hicks, Antoine
  Rousseau, Frederic Thevenet, Ahmed Khacef, and Maria Carreon.
\newblock The 2020 plasma catalysis roadmap.
\newblock {\em Journal of Physics D: Applied Physics}, 53(44):443001, aug 2020.

\bibitem{doi:10.1063/1.3366240}
Thomas~C. Killian and Steven~L. Rolston.
\newblock Ultracold neutral plasmas.
\newblock {\em Physics Today}, 63(3):46--51, 2010.

\bibitem{Pohl_Pattard_Rost}
T.~Pohl, T.~Pattard, and J.~M. Rost.
\newblock Relaxation to nonequilibrium in expanding ultracold neutral plasmas.
\newblock {\em Phys. Rev. Lett.}, 94:205003, May 2005.

\bibitem{DIH_UCNP}
D.O. Gericke and M.S. Murillo.
\newblock Disorder-induced heating of ultracold plasmas.
\newblock {\em Contributions to Plasma Physics}, 43(5-6):298--301, 2003.

\bibitem{Kuzmin_ONeil}
S.~G. Kuzmin and T.~M. O’Neil.
\newblock Numerical simulation of ultracold plasmas.
\newblock {\em Physics of Plasmas}, 9(9):3743--3751, 2002.

\bibitem{Pai_2010}
David~Z Pai, Deanna~A Lacoste, and Christophe~O Laux.
\newblock Nanosecond repetitively pulsed discharges in air at atmospheric
  pressure{\textemdash}the spark regime.
\newblock {\em Plasma Sources Science and Technology}, 19(6):065015, nov 2010.

\bibitem{Rusterholtz_2013}
D~L Rusterholtz, D~A Lacoste, G~D Stancu, D~Z Pai, and C~O Laux.
\newblock Ultrafast heating and oxygen dissociation in atmospheric pressure air
  by nanosecond repetitively pulsed discharges.
\newblock {\em Journal of Physics D: Applied Physics}, 46(46):464010, oct 2013.

\bibitem{Popov_2011}
N~A Popov.
\newblock Fast gas heating in a nitrogen{\textendash}oxygen discharge plasma:
  I. kinetic mechanism.
\newblock {\em Journal of Physics D: Applied Physics}, 44(28):285201, jun 2011.

\bibitem{Popov_2016}
N~A Popov.
\newblock Pulsed nanosecond discharge in air at high specific deposited energy:
  fast gas heating and active particle production.
\newblock {\em Plasma Sources Science and Technology}, 25(4):044003, may 2016.

\bibitem{Ono_2008}
Ryo Ono and Tetsuji Oda.
\newblock Measurement of gas temperature and {OH} density in the afterglow of
  pulsed positive corona discharge.
\newblock {\em Journal of Physics D: Applied Physics}, 41(3):035204, jan 2008.

\bibitem{Mintoussov_2011}
E~I Mintoussov, S~J Pendleton, F~G Gerbault, N~A Popov, and S~M Starikovskaia.
\newblock Fast gas heating in nitrogen{\textendash}oxygen discharge plasma:
  {II}. energy exchange in the afterglow of a volume nanosecond discharge at
  moderate pressures.
\newblock {\em Journal of Physics D: Applied Physics}, 44(28):285202, jun 2011.

\bibitem{HofmannPSST2011}
S~Hofmann, A~F~H van Gessel, T~Verreycken, and P~Bruggeman.
\newblock Power dissipation, gas temperatures and electron densities of cold
  atmospheric pressure helium and argon {RF} plasma jets.
\newblock {\em Plasma Sources Science and Technology}, 20(6):065010, nov 2011.

\bibitem{vanderHorst2012}
R~M van~der Horst, T~Verreycken, E~M van Veldhuizen, and P~J Bruggeman.
\newblock Time-resolved optical emission spectroscopy of nanosecond pulsed
  discharges in atmospheric-pressure n2 and n2/h2o mixtures.
\newblock {\em Journal of Physics D: Applied Physics}, 45(34):345201, aug 2012.

\bibitem{Adamovich2000}
Igor~V. Adamovich.
\newblock Three-dimensional analytic model of vibrational energy transfer in
  molecule-molecule collisions.
\newblock {\em AIAA Journal}, 39(10):1916--1925, 2001.

\bibitem{Guerra_2019}
Vasco Guerra, Antonio~Tejero del Caz, Carlos~D Pintassilgo, and Lu{\'{\i}}s~L
  Alves.
\newblock Modelling n2-o2 plasmas: volume and surface kinetics.
\newblock {\em Plasma Sources Science and Technology}, 28(7):073001, jul 2019.

\bibitem{DongAPL2005}
Lifang Dong, Junxia Ran, and Zhiguo Mao.
\newblock Direct measurement of electron density in microdischarge at
  atmospheric pressure by stark broadening.
\newblock {\em Applied Physics Letters}, 86(16):161501, 2005.

\bibitem{ParkevichPSST2019}
E~V Parkevich, M~A Medvedev, G~V Ivanenkov, A~I Khirianova, A~S Selyukov, A~V
  Agafonov, Ph~A Korneev, S~Y Gus'kov, and A~R Mingaleev.
\newblock Fast fine-scale spark filamentation and its effect on the spark
  resistance.
\newblock {\em Plasma Sources Science and Technology}, 28(9):095003, sep 2019.

\bibitem{MinesiPSST2020}
N~Minesi, S~Stepanyan, P~Mariotto, G~D Stancu, and C~O Laux.
\newblock Fully ionized nanosecond discharges in air: the thermal spark.
\newblock {\em Plasma Sources Science and Technology}, 29(8):085003, aug 2020.

\bibitem{BatallerPRL2016}
A.~Bataller, S.~Putterman, S.~Pree, and J.~Koulakis.
\newblock Observation of shell structure, electronic screening, and energetic
  limiting in sparks.
\newblock {\em Phys. Rev. Lett.}, 117:085001, Aug 2016.

\bibitem{BatallerPRL2014}
A.~Bataller, G.~R. Plateau, B.~Kappus, and S.~Putterman.
\newblock Blackbody emission from laser breakdown in high-pressure gases.
\newblock {\em Phys. Rev. Lett.}, 113:075001, Aug 2014.

\bibitem{BatallerOL2019}
Alexander Bataller, Alexandra Latshaw, John~P. Koulakis, and Seth Putterman.
\newblock Dynamics of strongly coupled two-component plasma via ultrafast
  spectroscopy.
\newblock {\em Opt. Lett.}, 44(23):5832--5835, Dec 2019.

\bibitem{Verreycken_2012}
T~Verreycken, R~M van~der Horst, A~H F~M Baede, E~M~Van Veldhuizen, and P~J
  Bruggeman.
\newblock Time and spatially resolved {LIF} of {OH} in a plasma filament in
  atmospheric pressure he-h2o.
\newblock {\em Journal of Physics D: Applied Physics}, 45(4):045205, jan 2012.

\bibitem{Heberlein_2009}
J~Heberlein, J~Mentel, and E~Pfender.
\newblock The anode region of electric arcs: a survey.
\newblock {\em Journal of Physics D: Applied Physics}, 43(2):023001, dec 2009.

\bibitem{doi:https://doi.org/10.1002/0471724254.ch3}
Michael~A. Lieberman and Allan~J. Lichtenberg.
\newblock {\em Atomic Collisions}, chapter~3, pages 43--85.
\newblock John Wiley `I\&' Sons, Ltd, 2005.

\bibitem{Scheiner_2020}
Brett Scheiner and Scott~D. Baalrud.
\newblock Mean force kinetic theory applied to self-diffusion in supercritical
  lennard-jones fluids.
\newblock {\em The Journal of Chemical Physics}, 152(17):174102, 2020.

\bibitem{Daligault_diffusion}
J\'er\^ome Daligault.
\newblock Diffusion in ionic mixtures across coupling regimes.
\newblock {\em Phys. Rev. Lett.}, 108:225004, May 2012.

\bibitem{LAMMPS}
A.~P. Thompson, H.~M. Aktulga, R.~Berger, D.~S. Bolintineanu, W.~M. Brown,
  P.~S. Crozier, P.~J. in~'t Veld, A.~Kohlmeyer, S.~G. Moore, T.~D. Nguyen,
  R.~Shan, M.~J. Stevens, J.~Tranchida, C.~Trott, and S.~J. Plimpton.
\newblock {LAMMPS} - a flexible simulation tool for particle-based materials
  modeling at the atomic, meso, and continuum scales.
\newblock {\em Comp. Phys. Comm.}, 271:108171, 2022.

\bibitem{BausPR1980}
Marc Baus and Jean-Pierre Hansen.
\newblock Statistical mechanics of simple coulomb systems.
\newblock {\em Physics Reports}, 59(1):1--94, 1980.

\bibitem{KILLIAN200777}
T.C. Killian, T.~Pattard, T.~Pohl, and J.M. Rost.
\newblock Ultracold neutral plasmas.
\newblock {\em Physics Reports}, 449(4):77--130, 2007.

\bibitem{KuzminPRL2002}
S.~G. Kuzmin and T.~M. O'Neil.
\newblock Numerical simulation of ultracold plasmas: How rapid intrinsic
  heating limits the development of correlation.
\newblock {\em Phys. Rev. Lett.}, 88:065003, Jan 2002.

\bibitem{TiwariPRE2017}
Sanat~Kumar Tiwari, Nathaniel~R. Shaffer, and Scott~D. Baalrud.
\newblock Thermodynamic state variables in quasiequilibrium ultracold neutral
  plasma.
\newblock {\em Phys. Rev. E}, 95:043204, Apr 2017.

\bibitem{MD_Frenkel}
Daan Frenkel and Berend Smit.
\newblock Chapter 6 - molecular dynamics in various ensembles.
\newblock In Daan Frenkel and Berend Smit, editors, {\em Understanding
  Molecular Simulation (Second Edition)}, pages 139--163. Academic Press, San
  Diego, second edition edition, 2002.

\bibitem{Ferziger}
J.~H. Ferziger and H.~G. Kaper.
\newblock Mathematical theory of transport processes in gases.
\newblock {\em American Journal of Physics}, 41(4):601--603, 1973.

\bibitem{HANSEN201313}
Jean-Pierre Hansen and Ian~R. McDonald.
\newblock Chapter 2 - statistical mechanics.
\newblock In Jean-Pierre Hansen and Ian~R. McDonald, editors, {\em Theory of
  Simple Liquids (Fourth Edition)}, pages 13--59. Academic Press, Oxford,
  fourth edition edition, 2013.

\bibitem{OttPOP2014}
T.~Ott, M.~Bonitz, L.~G. Stanton, and M.~S. Murillo.
\newblock Coupling strength in coulomb and yukawa one-component plasmas.
\newblock {\em Physics of Plasmas}, 21(11):113704, 2014.

\end{thebibliography}

\end{document}